\documentclass[11pt,a4paper,twocolumn]{article}
\usepackage[
left=2.0cm,
right=2.0cm,
top=2.5cm,
bottom=2.5cm
]{geometry}

\newcommand{\hi}{{\sc H\,i} }
\newcommand{\Mhi}{M$_{\text{\sc H\,i}}$\,}

\usepackage{hyperref}
\usepackage{xcolor}
\usepackage{natbib}
\usepackage{subfig}
\usepackage{array}
\usepackage{amsmath}
\usepackage{graphicx}
\usepackage{subfig}
\usepackage{cuted}
\usepackage{orcidlink}
\input{journals.def}

\title{A machine learning approach to estimating \hi deficiency in galaxies}
\date{}

\begin{document}

\twocolumn[

\maketitle
\author{
	Filip Jan\'ak\orcidlink{0009-0002-3680-672X}\textsuperscript{1,*} \and
	Boris Deshev\orcidlink{0000-0002-7898-5490}\textsuperscript{2} \and
	Roman Nagy\orcidlink{0000-0001-6327-3880}\textsuperscript{1} \and
	Rhys Taylor\orcidlink{0000-0002-3782-1457}\textsuperscript{3}\\[2ex]
	\textsuperscript{1}Faculty of Mathematics, Physics, and Informatics, Comenius University, Mlynsk\'{a} dolina, 842~48 Bratislava, Slovakia\\
	\textsuperscript{2}Tartu Observatory, University of Tartu, Observatooriumi~1, 61602 T\~oravere, Estonia\\
	\textsuperscript{3}Astronomical Institute of the Czech Academy of Sciences, Prague, Czechia\\[1ex]
	\textsuperscript{*}\texttt{filip.janak@fmph.uniba.sk}
}

\begin{abstract}

Measurements of the \hi content of galaxies serve as an important tracer for probing the impact of environment on galaxy evolution. More specifically, the \hi deficiency (defined as the difference between expected \textit{unaltered} and observed \hi content of a galaxy) is closely related with environmental effects, which are most significant in large groups and clusters. In this work, we aim to estimate the \hi deficiency of ALFALFA galaxies and investigate its relation with galactic environment. Using a random forest machine learning algorithm, we developed a predictive model capable of estimating the original \hi content of a galaxy based solely on its optical properties. The model was trained on a subsample of 6\,982 isolated ALFALFA galaxies with optical photometric data from the Sloan Digital Sky Survey (SDSS). Our predictive model outperforms the traditional approach, in which \hi mass is linearly related to optical size (both on a logarithmic scale). The model achieves $ \mathrm{RMSE} \approx 0.22 $~dex and $ R^2 \approx 0.80 $, compared with $ \mathrm{RMSE} \approx 0.26 $~dex and $ R^2 \approx 0.70 $ for the traditional method. We applied this model to predict the expected \hi content for non-isolated ALFALFA galaxies, enabling the calculation of \hi deficiency. Controlling for the effects of internal factors, like stellar mass and presence of AGN, we find an increase in binned median \hi deficiency of 0.15 dex attributable to environmental effects. In addition, we evaluate the temporal evolution of the predicted \hi mass, and associated \hi deficiency, due to the evolving stellar populations, following a gas removal event.

\end{abstract}

\textbf{Keywords (UAT):} Galaxies (573); Interstellar atomic gas (833)

\vspace{1em}

]

\section{Introduction}
\label{ch1}

Neutral atomic hydrogen ({\sc H\,i}) constitutes an important part of the interstellar medium as it is ultimately the fuel for star formation and thus a crucial part of the baryonic cycle \citep{2013ApJ...772..119L}. Its detection relies on the forbidden 21 cm hydrogen emission line, which is located in the radio part of the electromagnetic spectrum. Unlike parameters observed in the visible part of the spectrum, the \hi content of galaxies is a more challenging quantity to detect. Because of the intrinsic weakness of the line, a census of \hi content for a statistically significant sample of galaxies is currently available only at redshifts $z\ll 0.2$ \citep{2020MNRAS.496.3531G}. A comprehensive overview of the cold interstellar medium in the local universe is provided e.g. by \cite{2022ARA&A..60..319S}.

In this work, we define the \hi content of a galaxy as the decimal logarithm of its total \hi mass in the solar mass units (\Mhi), which is calculated with the standard formula \citep[see][]{2018ApJ...861...49H}

\begin{equation}
	M_{\mathrm{H\,I}}=2.356 \times 10^5 D^2 S_{21},
	\label{MHI}
\end{equation}

where $ S_{21} $ is the integrated \hi line flux density of a galaxy in $\mathrm{Jy}~\mathrm{km}~\mathrm{s}^{-1}$, and $ D $ is a distance in $\mathrm{Mpc}$. 

In addition to internal factors (e.g. presence or absence of an active galactic nucleus (AGN), star formation rate, total mass) the \hi content of a galaxy can be dramatically altered by various external processes. These include mechanisms such as tidal interactions among galaxies or between galaxies and the cluster, galaxy harassment, ram-pressure stripping (RPS), viscous stripping, thermal evaporation and starvation (see \citealt{2006PASP..118..517B} for more details). Understanding which of these processes dominate requires knowledge of the \textit{galaxy environment}, i.e. the intergalactic density around a given galaxy, with the dense environment corresponding to a galaxy being located within a large group or cluster. However, there are various tracers of the galactic environment (see Sect. \ref{ch2.3}), which can be sensitive to different physical mechanisms \citep{2017A&A...602A.100T}. For example, the distance to the $n$-th nearest neighbor is more sensitive to tidal effects and mergers whereas the luminosity density traces environment where starvation is likely to dominate. On the other hand, the friends-of-friends clustering \citep{2016A&A...588A..14T} identifies galaxy membership in larger structures \citep{1976ApJS...32..409T,1982ApJ...257..423H,1982Natur.300..407Z} and thus its susceptibility to RPS, galaxy harassment, starvation and lack of gas accretion.

Since the environmental effects on the \hi content of galaxies are complex, it is more instructive to first study galaxies that do not experience them. We refer to such galaxies as \textit{isolated} (see Sect. \ref{ch3.2}). These galaxies typically contain a gaseous disk which is larger than its optical disk \citep{2006PASP..118..517B}. Considering only isolated galaxies (IG), we can establish the expected (unaltered) \hi content and investigate how much environmental processes affect it (see Sect. \ref{ch4}). This, in return, enables us to study the role of environment in shaping galaxies. 

This approach was pioneered by \cite{1984AJ.....89..758H}, who defined a standard \hi content for IG and developed a predictive model based on optical properties such as total luminosity, optical diameter and morphological type. Using a sample of around three hundred field galaxies precisely observed in the 21 cm line with the Arecibo telescope, they examined several parameters which could be used to predict the expected \hi content such as the total luminosity and optical diameter. Their results showed that \Mhi scales linearly with optical diameter $D$ (both on the logarithmic scale) as

\begin{equation}
	\mathrm{log}_{10} \left( M^{exp}_{\mathrm{H\,I}} \right) = a + b \cdot \mathrm{log}_{10}(D),
	\label{MHI_vs_D}
\end{equation}

for parameters $a$ and $b$, weakly dependent on morphology. They also found that \Mhi divided by the squared \hi diameter (i.e. \hi surface density averaged over the \hi disk) is nearly constant for all IG. For further discussion of the \hi size-mass relation, see \cite{2016MNRAS.460.2143W} and \cite{2019MNRAS.490...96S}.

\cite{1984AJ.....89..758H} also established the formal definition of the \hi deficiency as the difference between logarithms of the expected and observed \Mhi of a galaxy 

\begin{equation}
	\mathrm{DEF}_{\mathrm{H\,I}}=\mathrm{log}_{10} \left( \dfrac{M^{exp}_{\mathrm{H\,I}}}{M_\odot} \right) - \mathrm{log}_{10} \left( \dfrac{M^{obs}_{\mathrm{H\,I}}}{M_\odot} \right),
	\label{HIdef}
\end{equation}

where the expected \hi mass represents the \Mhi of a galaxy if it were isolated. Thus a galaxy with deficiency $\mathrm{DEF}_{\mathrm{H\,I}}=+1.0$ has only 10\% of the \hi content of a comparable isolated galaxy, and one with $\mathrm{DEF}_{\mathrm{H\,I}}=+2.0$ would have only 1\%, etc. The \hi deficiency has been investigated in a number of studies, exploring dense galactic group \citep{1997A&A...325..473H,2009MNRAS.400.1962K}, galaxies in the Virgo cluster \citep{1983AJ.....88..881G,1985A&A...151..108G,2010A&A...518L..49C}, influence of the galactic environment \citep{2001ApJ...548...97S,2009MNRAS.400.1962K,2016MNRAS.455.1294D,2020MNRAS.499.3233R}, as well as the relation between the neutral and molecular hydrogen content of galaxies \citep{1991A&A...249..359C}. With the GALEX Arecibo SDSS Survey (GASS), \cite{2010MNRAS.403..683C} analyzed the \hi scaling relations, later expanded by the extended GALEX Arecibo SDSS Survey (xGASS) \citep{2018MNRAS.476..875C}. The environmental effects on GASS galaxies were investigated by \cite{2013MNRAS.436...34C}. 

Many of the studies cited above have determined the coefficients $a$ and $b$ using their own samples \citep[see e.g.][]{1984AJ.....89..758H,1996ApJ...461..609S,2011ApJ...732...93T,2014MNRAS.444..667D,2018A_A...609A..17J}. Most of them typically find an intrinsic scatter in the \hi deficiency around 0.3-0.4, meaning that while \hi deficiency can be estimated for individual galaxies, it only broadly constrains the actual gas loss. Generally, galaxies with $\mathrm{DEF}_{\mathrm{H\,I}} > 0.6$ are considered strongly deficient, $0.3 < \mathrm{DEF}_{\mathrm{H\,I}} < 0.6$ moderately deficient, and $\mathrm{DEF}_{\mathrm{H\,I}} < 0.3$ would be regarded as non-deficient.

Reducing the scatter in \hi deficiency has the potential to give a more detailed insight into the gas-loss processes at work in different environments. For example, galaxies in rich clusters have been detected with deficiencies as high as $\mathrm{DEF}_{\mathrm{H\,I}} \approx 2$ \citep[e.g.][]{2012MNRAS.423..787T,2018A_A...609A..17J}, but in other environments deficiencies can be much more modest. \cite{2018ApJ...852..142C} attempted to use a larger sample to achieve statistically significant results even at \hi deficiencies below $0.2$. While deficiencies in large clusters may be high, in other environments gas loss may well occur at such a low level that measuring it on a per-galaxy basis becomes extremely difficult, as indeed the results of \cite{2018ApJ...852..142C} suggest.

Recent progress in this area has been made by \cite{2018A_A...609A..17J}. Using an AMIGA subsample of highly isolated galaxies with \hi detections and constraints (544 objects), they estimated $a$ and $b$ coefficients which reduced the scatter in \hi deficiency to $0.25$. They also found that \Mhi could be predicted equally well on the basis of optical magnitude rather than diameter. The improvement achieved by \cite{2018A_A...609A..17J} results from two key factors: the use of a highly isolated control sample (significantly more isolated than those in previous studies) and a regression fitting method that robustly incorporates upper limits for non-detections, which came from a variety of sources, although the authors carefully controlled for the different observational effects. 

In this work, we explore a different approach that builds upon two recent advances: we exploit the extensive galaxy catalog of \cite{2017A&A...602A.100T}, which enables the identification of isolated systems across a substantially larger sample, and we make use of the large, homogeneous and statistically robust sample of \hi detections from the ALFALFA survey. Remarkably, this resource has not yet been employed in similar analysis. While ALFALFA has dramatically expanded the number of \hi detected galaxies, most of its detections are spatially unresolved. This motivated us to use a new methodology for predicting the \hi mass from optical properties: the use of machine learning (ML) techniques to examine the correlation with \Mhi based on a much wider set of optical parameters than magnitude or diameter. The use of ML is gradually increasing in popularity, having a wide range of applications in astrophysics. \cite{2018MNRAS.479.4509R} used several ML models trained on the cosmological hydrodynamic simulation data to predict the neutral hydrogen content of galaxies. Based on the ALFALFA \hi survey, \cite{2017MNRAS.464.3796T} used artificial neural networks to predict the \hi gas mass fraction (i.e. the ratio between \Mhi and the stellar mass) for the SDSS galaxies. Similarly, \cite{2020ApJ...900..142W} used convolutional neural networks to predict the \hi gas mass fraction from SDSS images in g, r and i bands.

This paper is organized as follows: Sect. \ref{ch2} introduces the data used in our analysis. Sect. \ref{ch3} presents the classical predictive model as well as the ML random forest (RF) predictive model. In Sect. \ref{ch4} we apply the RF model to calculate the \hi deficiency and explore its relation with the galactic environment. We discuss and summarize our results in Sect. \ref{ch5}.

\section{Data}
\label{ch2}

In this Section, we briefly introduce the data used in this study: a reliable sample of galaxies detected in \hi (Sect. \ref{ch2.1}), their optical properties (Sect. \ref{ch2.2}) and information about their environment (Sect. \ref{ch2.3}). In Sect. \ref{ch2.4}, we explore the relation between \Mhi and galactic morphology, together with the optical parameters used as proxies for the morphology.

\subsection{\hi data}
\label{ch2.1}

The Arecibo Legacy Fast ALFA (ALFALFA) survey \citep{2005AJ....130.2598G} was a large drift scan survey carried out at Arecibo. It covered approximately 7\,000 square degrees of the sky to a root mean square noise level of 2.3~mJy at a velocity resolution of 10~km/s up to redshift $ z<0.06 $. With a beam size of $3.5'$ it had a column density sensitivity of around $10^{18}$  $\mathrm{atoms}~\mathrm{cm}^{-2}$. ALFALFA provides a large, homogeneous data set of approximately 31\,500 galaxies \citep{2018ApJ...861...49H}, making it an excellent source of \hi information for predicting the gas content of isolated galaxies. We used this catalog as the source of \hi data for the present work.

Based on the \hi line observation, this catalog provides the heliocentric velocity of the 21 cm line in the observed frame, the velocity width of the line at the $ 50\% $ and $ 20\% $ levels, the integrated \hi line flux and the logarithm of \Mhi (see Eq. \ref{MHI}), which is the subject of this work.

Although ALFALFA is among the most extensive and homogeneous extragalactic \hi surveys to date, it is not without certain observational limitations. Like all flux-limited surveys, ALFALFA is affected by the Malmquist bias, which favors intrinsically brighter (more \hi rich) galaxies at larger distances. The survey completeness depends on both the total \hi flux and the velocity width of the line, as quantified by \citep{2011AJ....142..170H}, who provide completeness curves at different confidence levels. More recently, \citep{2025A&A...696A.113T} recalculated the completeness relation based on the integrated signal-to-noise ratio (SNR), which offers a convenient way to assess the detectability of sources with given line properties. Their results suggest that ALFALFA detections become increasingly complete at high SNR. Therefore, we restrict our analysis to galaxies with confirmed \hi detections and do not include upper limits for non-detections. This means our sample naturally favors late-type galaxies (see Sect. \ref{ch2.4} and Fig. \ref{morph}) which tend to be gas rich. We discuss the implications of these limitations on our work in Sect. \ref{ch5}. 

\subsection{Optical data}
\label{ch2.2}

The Sloan Digital Sky Survey (SDSS) is one of the most influential optical surveys in astronomy, providing deep, multi-band imaging and spectroscopy over a large fraction of the sky \citep{2000AJ....120.1579Y}. Conducted with the 2.5-meter telescope at Apache Point Observatory, SDSS has mapped hundreds of millions of celestial objects in five optical bands ($u$, $g$, $r$, $i$, $z$), delivering precise photometric and astrometric measurements. 

\cite{2020AJ....160..271D} released the ALFALFA-SDSS catalog, which provides (along with several other derived astrophysical quantities) a cross-match between ALFALFA and the SDSS DR15 photometric catalog \citep{2017AJ....154...28B}. The cross-matching was performed using the SDSS Cross ID tool, searching within a radius of $0.1'$ around the optical counterpart positions identified in the ALFALFA database, with galaxies showing anomalous magnitudes inspected individually. This catalog served as the source of input (optical) data for our predictive model (see Sect. \ref{ch3}). 

From SDSS we primarily used Petrosian magnitudes, Petrosian radii, Petrosian radii containing $ 50\% $ and $ 90\% $ of the total flux (each of them in g, r and i bands). The Petrosian system \citep{1976ApJ...209L...1P}, as implemented in SDSS \citep{2001AJ....121.2358B}, provides a consistent approach to measure galaxy brightness and size that is relatively insensitive to distance and surface-brightness dimming. It is based on the Petrosian ratio, $r_P(r)$, defined as the ratio of the local surface brightness in an annulus at radius $r$ to the mean surface brightness within that radius. The Petrosian radius, $R_P$, is defined as the radius where $r_P(R_P)=0.2$. The Petrosian flux is then the total flux integrated within $2\cdot R_P$ and the Petrosian magnitude is the corresponding flux converted to the magnitude scale. The Petrosian half-light radius ($R_{50}$) and ninety-percent radius ($R_{90}$) correspond to the radii enclosing $ 50\% $ and $ 90\% $ of the Petrosian flux, respectively.

In addition, the catalog of \citep{2020AJ....160..271D} provides other useful parameters such as stellar masses. We adopt the stellar masses (calculated with the method of Taylor \citealt{2011MNRAS.418.1587T} from SDSS data) in Sect. \ref{ch4} to mitigate the influence of internal factors on \hi deficiency.

\subsection{Galactic environment}
\label{ch2.3}

There are many ways to quantify the galactic environment, e.g. \cite{2016A&A...588A..14T,2018A&A...618A..81T,2021ApJ...909..143Y}. Within SDSS, there are several works based on spectroscopy \citep{2012ApJS..199...34W,2012A&A...540A.106T,2017A&A...602A.100T} or photometry \citep{2014ApJ...785..104R}. In general, spectroscopic catalogs contain fewer sources, cover lower redshifts ($z < 0.2$), and provide more accurate distances, while photometric catalogs contain more sources at higher redshifts ($0.05 < z < 0.8$) but with less accurate distances. A precise estimate of the environment around galaxies requires spectroscopic redshifts. Moreover, the redshift range prevents us from using photometric catalogs in this work, since most ALFALFA galaxies have $z < 0.05$. Therefore, we adopt the spectroscopic catalog of \cite{2017A&A...602A.100T}, which offers the most complete cross-match with ALFALFA-SDSS, providing environmental information for approximately half of the galaxies (i.e. those with available spectroscopy). Since this catalog is constructed from SDSS spectroscopic targets, it is inherently biased toward more luminous galaxies, as spectroscopy is generally unavailable for fainter systems. A comparison between the used and discarded samples is presented in the Appendix \ref{Appendix_0}.

The environment catalog by \cite{2017A&A...602A.100T} was constructed using a modified friends-of-friends group finder algorithm applied to SDSS DR12 spectroscopic data. It comprises around 584\,000 galaxies among which authors identified about 88\,000 groups. From this catalog, we used four environmental tracers: the number of galaxies $ N_{gal} $ in a group ($ N_{gal} = 1 $ indicates an isolated galaxy), the estimated mass of the group a galaxy belongs to $ M_{200} $ (mass located within the sphere where the mean density is 200 times larger than the mean density of the universe) and normalized environmental densities smoothed on scales of 1.5 and 3 Mpc.

\subsection{Galactic morphology}
\label{ch2.4}

It is natural to incorporate morphology into our analysis as it is expected to relate to the total \hi mass. Early-type galaxies (ETGs) are usually \hi depleted whereas late-type galaxies (LTGs) tend to be \hi rich, often featuring significant star formation. Analyzing a few hundred nearby galaxies, \cite{2012MNRAS.422.1835S} found that ETGs generally contain less \hi gas than LTGs (the distribution of ETGs is broader and peaks at lower $ \mathrm{log}_{10}(M_{\mathrm{H\,I}}) $ than the distribution of LTGs), although they reported that the overlap between both distributions is significant. A similar trend was observed for the ratio of \Mhi to total luminosity, with ETGs showing a broader distribution that peaks at lower values. The authors further demonstrated that the substantial distinction between ETGs and LTGs lies in the column (surface) density: ETGs, even when containing similar \Mhi, exhibit much lower surface densities, which suppresses star formation.

Apart from visual identification, determining morphology is not straightforward. Nevertheless, there are several measurable properties that correlate with the morphological type. Using SDSS data, \cite{2001AJ....122.1861S} and \cite{2001AJ....122.1238S} demonstrated that important morphology tracers are colors (ETGs are generally redder, LTGs bluer) and surface-brightness radial profiles. The radial surface-brightness profile of a galaxy is commonly described with the Sérsic profile \citep{1963BAAA....6...41S} parameterized by index $ n $. For majority of galaxies, the Sérsic index lies between $ n \in (0.5,10) $. For $ n = 1 $, the Sérsic profile reduces to the exponential profile which is typical for spiral disks, while $ n = 4 $ corresponds with the de Vaucouleurs law \citep{1948AnAp...11..247D} that reasonably describes ETGs and bulges. 

Although the SDSS does not directly compute Sérsic indices or full surface-brightness profiles, it provides several useful proxies. Specifically, it reports the likelihoods of exponential and de Vaucouleurs fits (probabilities of achieving the measured chi-squared for the exponential and de Vaucouleurs fits, respectively), which can serve as indicators of morphological type. If the exponential fit of a given galaxy has larger likelihood than the de Vaucouleurs fit, it is more likely a LTG than an ETG and vice versa. In addition, the SDSS provides Petrosian radii containing $ 50\% $ and $ 90\% $ of the total flux, which allow the computation of the concentration index, another useful tracer of the surface-brightness profile and morphology. Concentration index is defined as the ratio of $R_{90}$ and $R_{50}$ (in a given band)

\begin{equation}
	c_{x}=\frac{R_{90,x}}{R_{50,x}}.
	\label{concentration}
\end{equation}

Generally, LTGs have lower concentration index ($ c \approx 2.3 $ for exponential profile) than ETGs ($ c \approx 5.5 $ for de Vaucouleurs law). 

\begin{figure*}
	\centering
	\subfloat{\includegraphics[width=0.95\textwidth]{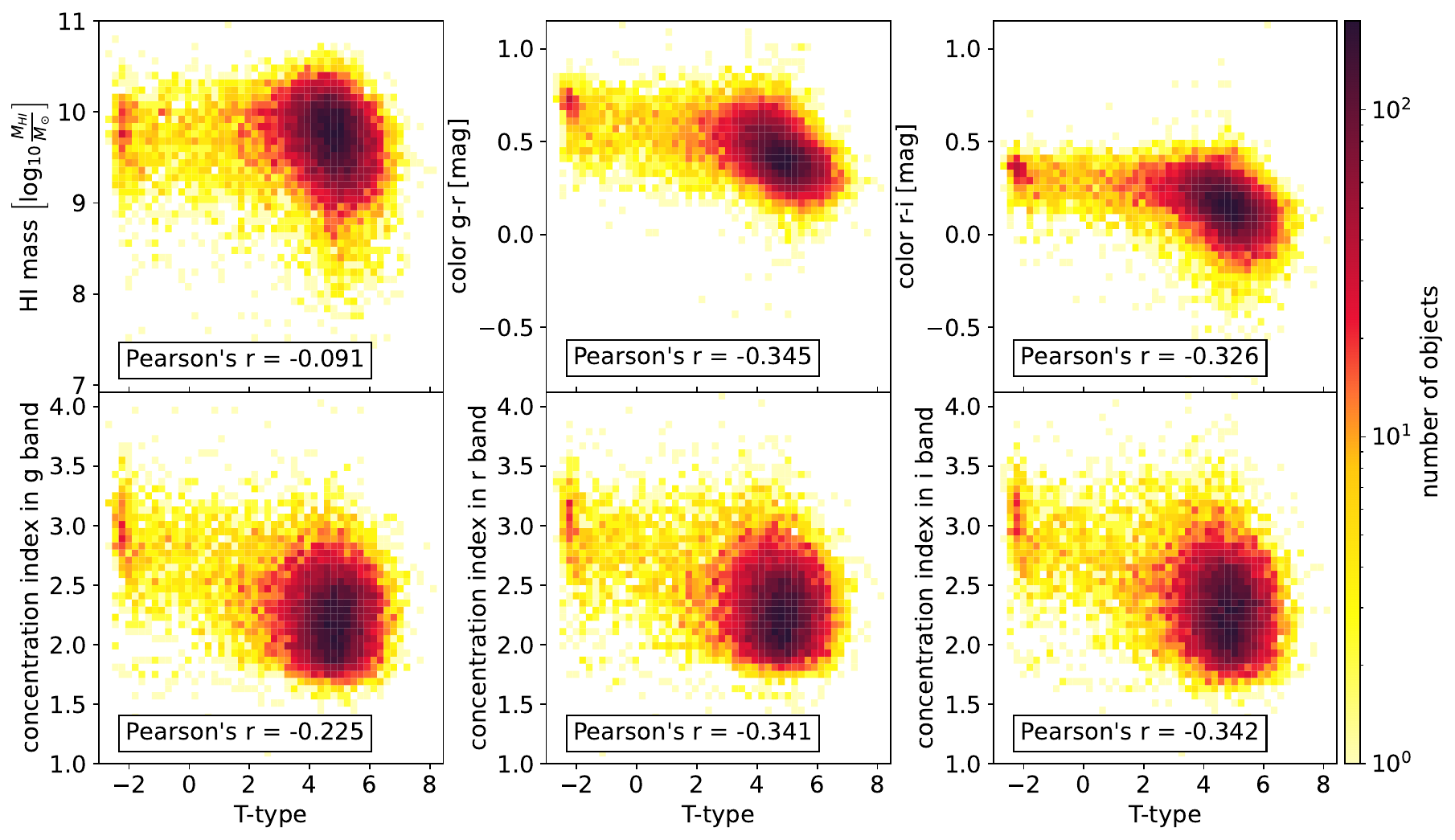}}
	\caption{The morphology distribution of 14\,288 ALFALFA sources with respect to the \Mhi, colors and concentration indices. For each plot the Pearson correlation coefficient is given.}
	\label{morph}
\end{figure*}

The most reliable method of morphology classification is visual inspection. Galaxy Zoo 1 (GZ1) \citep{2008MNRAS.389.1179L} used help from volunteers for morphological classification of almost 900\,000 galaxies. Its successor, the Galaxy Zoo 2 (GZ2) \citep{2013MNRAS.435.2835W} featured much more detailed morphology classification for approximately 240\,000 brightest galaxies. To demonstrate the relation between optical morphology tracers (color and concentration indices) and the morphological type, we used the SDSS morphology catalog created by \cite{2018MNRAS.476.3661D}, which presents T-type and GZ2-like classification for around 670\,000 galaxies. This catalog was built using deep learning algorithms with convolutional neural networks trained on GZ2 and the morphology catalog of \cite{2010ApJS..186..427N}. For our purposes, we used the T-type morphology classification, which quantifies the relative importance between the disk and the bulge \citep{1963ApJS....8...31D}. T-type values are negative for ETGs and positive for LTGs. The exact T values for each morphology sub-type are given in Table \ref{ttype_tab} (as listed in Table 1. by \citealt{2010ApJS..186..427N}, with minor modifications by \citealt{2018MNRAS.476.3661D}). It should be noted that the non-integer T values in the \cite{2018MNRAS.476.3661D} catalog arise from the regression-based approach adopted in their predictive model. After cross-matching with the ALFALFA-SDSS, approximately one half of the sources remained. Their morphology distribution with respect to \Mhi, colors and concentration indices is shown in Fig. \ref{morph}.

\begin{table}
	\centering
	\caption{T-Type classification scheme.}
	\begin{tabular}{cc}
		\hline
		\textbf{Class} & \textbf{T-type}\\
		\hline\hline
		C0 & -3  \\
		\hline
		E0 & -3  \\
		\hline
		E+ & -3  \\
		\hline
		S0- & -2  \\
		\hline
		S0 & -1  \\
		\hline
		S0+ & -1  \\
		\hline
		S0/a & 0  \\
		\hline
		Sa & 1  \\
		\hline
		Sab & 2  \\
		\hline
		Sb & 3  \\
		\hline
		Sbc & 4  \\
		\hline
		Sc & 5  \\
		\hline
		Scd & 6  \\
		\hline
		Sd & 7  \\
		\hline
		Sdm & 8  \\
		\hline
		Sm & 9  \\
		\hline
		Im & 10  \\
		\hline
	\end{tabular}
	\label{ttype_tab}
\end{table}

The upper left panel of Fig. \ref{morph} shows no clear correlation between \Mhi and the morphology type in our sample. In addition to mechanisms discussed in the first paragraph of this Section, there may be a role of an observational bias caused by the \hi measurement sensitivity. In ETGs, the \hi component - when present - is often kinematically disturbed or distributed in extended, low surface density structures. This reduces the detectability of their \hi emission by lowering the SNR, making it more challenging to detect weak \hi signals spread over a broad velocity range. Moreover, there is a selection effect for dwarf galaxies in play. In clusters, dwarf ellipticals are expected to be \hi-poor due to RPS, while in the field, such objects are rare \citep{2003AJ....125.2975L,2016MNRAS.455.2323M,2020MNRAS.494.1114S}. Thus it is reasonable to conclude that the low \Mhi tail corresponds to dwarf LTGs, while low \Mhi ETGs are underrepresented in our sample.

The other panels of Fig. \ref{morph} show T-type distributions with respect to optical morphology tracers (colors and concentration indices). In our sample, all of these tracers are related with the morphology exposing weak negative correlation (see Pearson's r in Fig. \ref{morph}), although the scatter is significant. The $ \mathrm{g} - \mathrm{r} $ color expresses the strongest correlation, followed by the concentration indices in $ \mathrm{i} $ and $ \mathrm{r} $ bands. Overall, it is evident from all panels in Fig. \ref{morph} that the distribution of our sample is heavily shifted towards LTGs (this is discussed further in Sect. \ref{ch5}).

\section{Predictive model}
\label{ch3}

In this Section, we describe in detail our predictive model for estimating \Mhi from optical data. We first explain how we processed the data that were used to create the model (Sect. \ref{ch3.1}), then introduce our final sample of IG (Sect. \ref{ch3.2}). We proceed with the linear predictive model (Sect. \ref{ch3.3}) and finish with the RF model (Sect. \ref{ch3.4}).

To evaluate the model performance, we use standard regression metrics. The model predicts the desired \textit{target variable} from a set of input variables called \textit{features} (predictors). When creating a predictive model, the data set is usually split into the training and testing subsets. The model learns only on the training data, and the test set is subsequently used to assess its performance on unseen data. Let us consider a model $ f $ trained on the training set. The mean squared error (MSE) of the model $ f $ evaluated on the test set consisting of features $ X $ and target variables $ y $ is 

\begin{equation}
	\mathrm{MSE}(f,X,y)=\frac{1}{n}\sum_{i=1}^{n}(\tilde{y}_i-y_i)^2,
	\label{mse}
\end{equation}

where $ \tilde{y}_i $ is the i-th predicted value of the model ($ \tilde{y} = f(X) $) and $ n $ is the number of sources in the test set. When discussing the performance of our model we quote RMSE, the square root of the MSE. The $ R^2 $ score is then defined as 

\begin{equation}
	R^2(f,X,y)=1-\frac{\mathrm{MSE}(f,X,y)}{\mathrm{MSE}(f_0,X,y)}.
	\label{r2}
\end{equation}

Here, $ f_0 $ denotes the baseline model, defined as the mean value of the target variable from the test set. A perfectly fitting model has $ R^2=1 $, whereas a model performing no better than the baseline model yields $ R^2=0 $. Worse models reach negative $ R^2 $ values.

\subsection{Data processing}
\label{ch3.1}

We began with the dataset described in Sect. \ref{ch2}.The absolute Petrosian magnitudes $ M_x $ ($ x $ = g, r, i bands) were calculated using the Pogson equation

\begin{equation}
	M_x = m_x - E_x - 5 \cdot \left( 5 + \mathrm{log}_{10}\left( \dfrac{d}{\mathrm{Mpc}} \right) \right) - K_x,
	\label{pogson}
\end{equation}

where $ m_x - E_x $ is the apparent magnitude in the $ x $ band corrected for the Galactic extinction (values adopted from the SDSS catalog, with no internal extinction correction applied), $ K_x $ is the K-correction \citep{2007AJ....133..734B} in the $ x $ band and $ d $ is the distance in Mpc from the ALFALFA catalog (see \citealt{2018ApJ...861...49H} for a detailed description of the distance determination). The color indices were computed as differences between corresponding absolute Petrosian magnitudes.

The apparent size of the galaxies $ R_{ap} $, given by SDSS in angular units, was converted to physical size $ R_{phys} $ in kpc with

\begin{equation}
	\frac{R_{phys}}{\mathrm{kpc}} = \mathrm{tan}\left( \frac{\pi \cdot R_{ap}}{\mathrm{arcsec} \cdot 1\,296\,000} \right) \dfrac{2\,000 \cdot d}{\mathrm{Mpc}}.
	\label{radius}
\end{equation}

Concentration indices in g, r, i bands were calculated with Eq. \ref{concentration}.

We do not consider any cosmological effects in this work, since our sample consists of relatively close objects ($ z < 0.06 $, or equivalently $ d \leq 260 $ Mpc). For the most distant galaxy in our sample, the physical radius obtained using Eq. \ref{radius} differs from that derived with the $ \mathrm{\Lambda CDM} $ cosmology model by only around $ 2\% $ which is negligible for our purposes.

\subsection{Sample of isolated galaxies}
\label{ch3.2}

To construct a model capable of predicting the expected \Mhi, we assembled a sample of isolated galaxies. Each galaxy in our sample is characterized by 17 optical features (which serve as predictors) and one target variable - the decadic logarithm of the total \hi mass (see Table \ref{samp_tab}). The IG sample consists of ALFALFA-SDSS galaxies classified as isolated. Numerous definitions of galaxy isolation have been proposed in the literature \citep{2011AstBu..66....1K,2011MNRAS.417..370G,2012MNRAS.424.2574W,2018A_A...609A..17J}, typically based on criteria requiring the absence of any significant companion within a specified distance or magnitude range. Here, we define isolated galaxies as those with $N_{gal}=1$ \citep{2017A&A...602A.100T}. To minimize the influence of internal processes, we excluded from the IG sample all galaxies hosting AGN, since AGN can significantly affect the \hi content of a galaxy. The presence of AGN, provided by SDSS spectroscopic catalog, is based on whether the galaxy has detectable emission lines that are consistent with being a Seyfert or low-ionization nuclear emission-line region (LINER) by the dividing line applied in the Baldwin–Philips–Terlevich (BPT) diagram \citep{1981PASP...93....5B}.

\begin{table*}
	\centering
	\caption{0.1th and 99.9th percentiles of features and the target variable of our sample of IG.}
	\begin{tabular}{ccccc}
		\hline
		\textbf{Type} & \textbf{Quantity} & \textbf{Unit} & \textbf{0.1th percentile} & \textbf{99.9th percentile}\\
		\hline\hline
		Feature & g & [mag] & -21.92 & -12.89 \\
		\hline
		Feature & r & [mag] & -22.57 & -13.13 \\
		\hline
		Feature & i & [mag] & -22.86 & -12.89 \\
		\hline
		Feature & $R_g$ & [kpc] & 0.46 & 39.41 \\
		\hline
		Feature & $R_r$ & [kpc] & 0.51 & 27.06 \\
		\hline
		Feature & $R_i$ & [kpc] & 0.53 & 47.60 \\
		\hline
		Feature & $R_{50,g}$ & [kpc] & 0.22 & 11.68 \\
		\hline
		Feature & $R_{50,r}$ & [kpc] & 0.24 & 10.72 \\
		\hline
		Feature & $R_{50,i}$ & [kpc] & 0.24 & 10.39 \\
		\hline
		Feature & $R_{90,g}$ & [kpc] & 0.48 & 26.84 \\
		\hline
		Feature & $R_{90,r}$ & [kpc] & 0.53 & 26.11 \\
		\hline
		Feature & $R_{90,i}$ & [kpc] & 0.49 & 26.28 \\
		\hline
		Feature & color g - r & [mag] & -0.03 & 1.17 \\
		\hline
		Feature & color r - i & [mag] & -0.77 & 0.64 \\
		\hline
		Feature & $c_g$ & [ - ] & 1.57 & 3.91 \\
		\hline
		Feature & $c_r$ & [ - ] & 1.62 & 3.84 \\
		\hline
		Feature & $c_i$ & [ - ] & 1.57 & 3.86 \\
		\hline
		Target variable & \Mhi & [$\mathrm{log}_{10}\frac{M_{\mathrm{H\,I}}}{M_\odot}$] & 7.13 & 10.58 \\
		\hline
	\end{tabular}
	\label{samp_tab}
\end{table*}

Finally, after applying these criteria: 
\begin{itemize}
	\item inclusion in the ALFALFA, ALFALFA–SDSS, and \cite{2017A&A...602A.100T} catalogs,
	\item isolation condition $N_{gal}=1$,
	\item exclusion of AGN
\end{itemize}
we obtained a final sample of 6\,982 isolated galaxies. Feature distributions of our sample with respect to the target variable are provided in the Appendix \ref{Appendix_1}. In Fig. \ref{corr_matrice} we show the correlation matrix for features and the target variable of IG sample. The first row (or column) shows the correlation of \Mhi with features. The strongest correlation is present for absolute magnitudes (g magnitude in particular). From all radii (which are also significantly correlated) the Petrosian radii with 90\% of the total flux yield the strongest relation. Across bands, the g band quantities performs better than quantities in other two bands. In contrast, colors show only weak correlations, and concentration indices are essentially uncorrelated.

\begin{figure}
	\centering
	\includegraphics[width=0.5\textwidth]{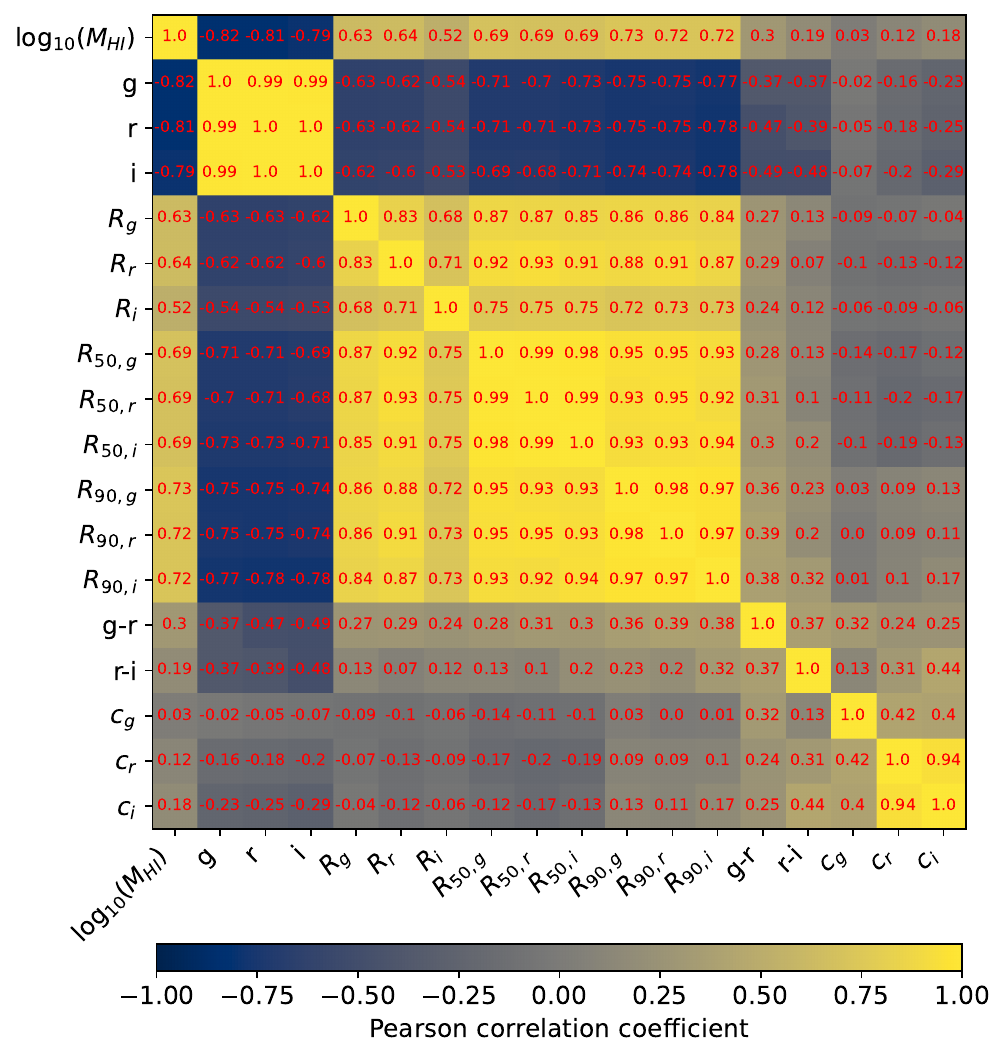}
	\caption{The Pearson correlation matrix for features and the target variable for our sample of IG.}
	\label{corr_matrice}
\end{figure}

\subsection{Classical model}
\label{ch3.3}

\begin{figure}
	\centering
	\includegraphics[width=0.5\textwidth]{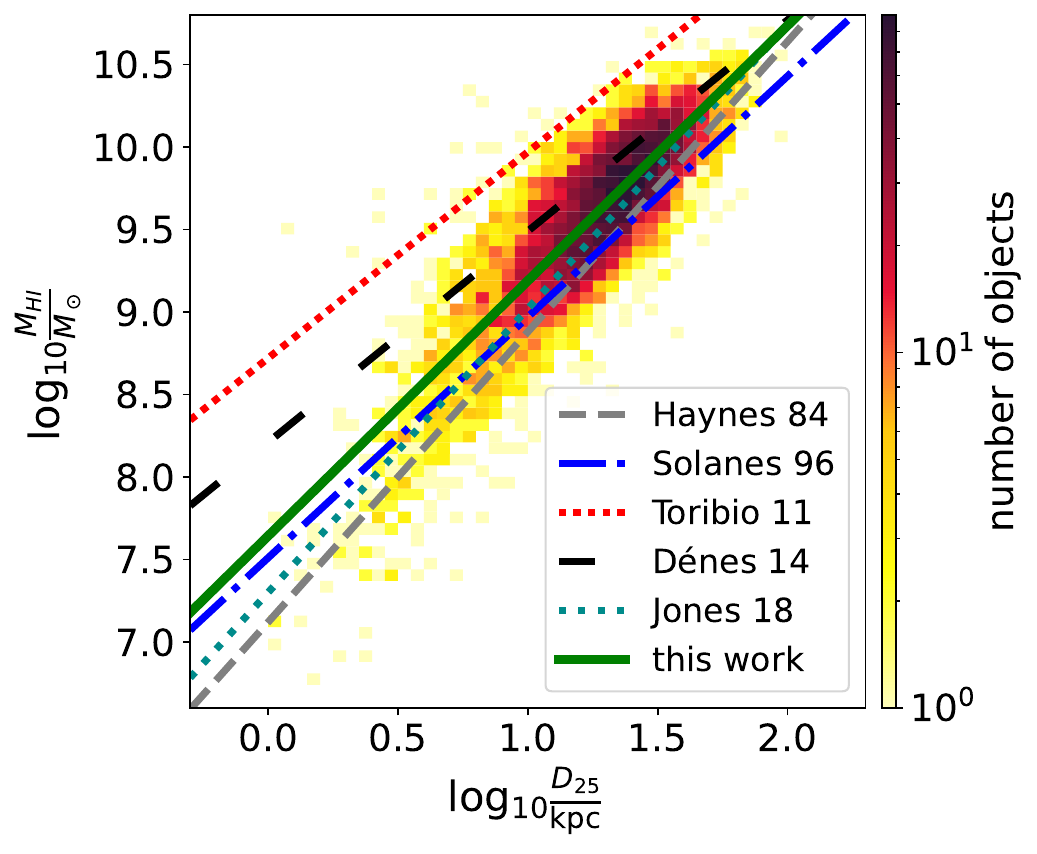}
	\caption{\hi mass from our sample of IG versus the optical diameter $ D_{25} $ linearly fitted with the green line, compared with \cite{1984AJ.....89..758H} (grey dashed line), \cite{1996ApJ...461..609S} (blue dashdotted line), \cite{2011ApJ...732...93T} (red dotted line), \cite{2014MNRAS.444..667D} (black loosely dashed line) and \cite{2018A_A...609A..17J} (cyan loosely dotted line).}
	\label{lin_model}
\end{figure}

The common approach in the literature relates \Mhi in IG to the optical diameter (both on the logarithmic scale) through a linear relation (see Eq. \ref{MHI_vs_D}). However, such results are difficult to compare directly, since the optical diameter can be defined in multiple ways and it depends on observational methods and data processing. For example, \cite{1984AJ.....89..758H} adopted blue band optical diameters from the UGC catalog, a system that was also used by \cite{1996ApJ...461..609S}. \cite{2014MNRAS.444..667D} and \cite{2018A_A...609A..17J} used optical diameters $ D_{25} $ (the diameter of a galaxy at the surface brightness of 25 mag arcsec$^{-2}$) in B band, and \cite{2011ApJ...732...93T} in the r band. In order to compare these results with our SDSS-based work, we adopted a correction inspired by \cite{2022A&A...665A.155D} to transform the optical radius of IG into $ D_{25} $. Specifically, we used the SDSS Petrosian radius containing 90\% of the total flux in g band, adopting the conversion $ D_{25} = 1.4 \times R_{90,g} $. This conversion was originally calibrated for gas-rich, late-type galaxies and is therefore expected to be most reliable for systems with similar T-type, which dominate our sample. Our results for the IG sample are presented in Fig. \ref{lin_model} and Table \ref{ab_tab}, where we compare derived linear fit with other studies. The established linear fit for our IG sample is given by Eq. \ref{lin_fit}

\begin{equation}
	\mathrm{log}_{10} \frac{M_{\mathrm{H\,I}}}{M_\odot} = (1.55 \pm 0.01)\times\mathrm{log}_{10}\frac{D_{25}}{\mathrm{kpc}} + (7.64 \pm 0.02).
	\label{lin_fit}
\end{equation}

\begin{table*}
	\centering
	\caption{Parameters $a$ and $b$ from Eq. \ref{MHI_vs_D}, describing the linear relation between \Mhi and optical diameter (both on a logarithmic scale) for isolated galaxies, compared with values reported in the literature.}
	\label{ab_tab}
	\begin{tabular}{cccc}
		\hline
		\textbf{Source} & \textbf{Optical diameter} & \textbf{a} & \textbf{b}\\
		\hline\hline
		\cite{1984AJ.....89..758H} & blue band (UGC) & 7.12 & 1.76  \\
		\hline
		\cite{1996ApJ...461..609S} & blue band (UGC) & $7.51^{+0.02}_{-0.03}$ & $1.46\pm0.02$  \\
		\hline
		\cite{2011ApJ...732...93T} & $D_{25}$ (r band) & $8.72\pm0.12$ & $1.25\pm0.13$  \\
		\hline
		\cite{2014MNRAS.444..667D} & $D_{25}$ (B band) & $8.21\pm0.04$ & $1.27\pm0.04$  \\
		\hline                        
		\cite{2018A_A...609A..17J} & $D_{25}$ (B band) & $7.30\pm0.12$ & $1.72\pm0.08$  \\
		\hline
		this work & $D_{25}$ (g band) & $7.64\pm0.02$ & $1.55\pm0.01$  \\
		\hline
	\end{tabular}
\end{table*}

We tested the performance of our linear model (Eq. \ref{lin_fit}) to find out its predicting capabilities on the IG sample, obtaining $\mathrm{RMSE} \approx 0.26~\mathrm{dex}$ and $R^2 \approx 0.70$. This level of accuracy is comparable to previous estimates of the expected \hi content of isolated galaxies. \cite{1984AJ.....89..758H} reported a standard error of about 0.20, while the scaling relations of \cite{2018A_A...609A..17J} predict the \hi mass to within 0.25 dex. A more detailed assessment of how our linear model performs in the context of \hi deficiency is presented in Sect. \ref{ch4}. We further compare the linear relation between \Mhi and the optical diameter for different morphologies in the Appendix \ref{Appendix_2}.

\subsection{Random forest model}
\label{ch3.4}

To develop a more advanced predictive model, we tested several supervised ML regression algorithms. We ultimately selected the RF algorithm \citep{2001MachL..45....5B} as it provides high accuracy for our problem while delivering good control and transparency of the created model. 

The RF algorithm is an ensemble learning method, i.e. it combines the outputs of many individual models (decision trees) to produce a final prediction. Each tree is trained on a bootstrapped subset (i.e. a random sample drawn with replacement from the original training set) and makes its own estimate of the target variable. This procedure ensures that individual trees are slightly different from each other, increasing the diversity of the ensemble. The final prediction is then obtained by averaging the predictions from all trees. This approach reduces overfitting and improves overall model robustness compared to using a single decision tree. During the construction of each decision tree, the algorithm splits the data based on feature values, choosing the split that minimizes the combined impurity of resulting subsets. Impurity measures how varied the target values are; lower impurity indicates greater homogeneity among the target values. In our model, we quantified the impurity using the MSE. A larger reduction in impurity indicates a more informative split. From these splits, the RF model can estimate intrinsic feature importances, indicating how strongly each input feature contributes to the final predictions. These intrinsic feature importances are based on the impurity decrease (quantified by the MSE, effectively corresponding to the variance reduction) within each tree and are available for the model itself. Intrinsic importances differ from permutation feature importances (computed by randomly shuffling the values of a single feature and measuring the resulting decrease in the model's performance) which require analysis on a test set. However, those two estimates of feature importances are not exactly the same. Intrinsic (impurity based) feature importances are biased towards high cardinality features (when a feature contains many unique values). 

Like most ML algorithms, the RF has two important limitations: it cannot extrapolate and it is unable to account for measurement errors (uncertainties). The inability to extrapolate, which emerges naturally from the tree-based construction of the RF, limits the application of our model to the range of physical properties sampled by the training set. The second limitation, if addressed, has a potential to improve model's performance and will be the subject of our future work. For example \cite{2019AJ....157...16R} created the probabilistic RF classification algorithm which naturally accounts for the errors in both features and target variables.

For our implementation, we optimized the hyperparameters (i.e. parameters that control the training process of the model) and adopted a configuration with 300 decision trees of unlimited depth, using 80\% of the original training set during the bootstrapping, considering 80\% of all features upon each split. We used the \textit{Scikit-learn} module \citep{2011JMLR...12.2825P}, which provides the RF implementation in the Python programming language. The RF model used the IG sample which was split into training and testing subsets in ratio 80:20. Compared to the linear model (Sect. \ref{ch3.3}) which used only one predictor, the RF model achieved a noticeably better performance, with $\mathrm{RMSE} \approx 0.22~\mathrm{dex}$ and $R^2 \approx 0.80$ on the test set. Although our feature set contains high number of predictors (17 features, see Table \ref{samp_tab}), with some of them being strongly correlated with each other (see Fig. \ref{corr_matrice}), dimensionality reduction techniques (such as principal component analysis) resulted in slightly worse predictive accuracy. Therefore we retained all 17 features in the final model.

\begin{figure}
	\centering
	\includegraphics[width=0.45\textwidth]{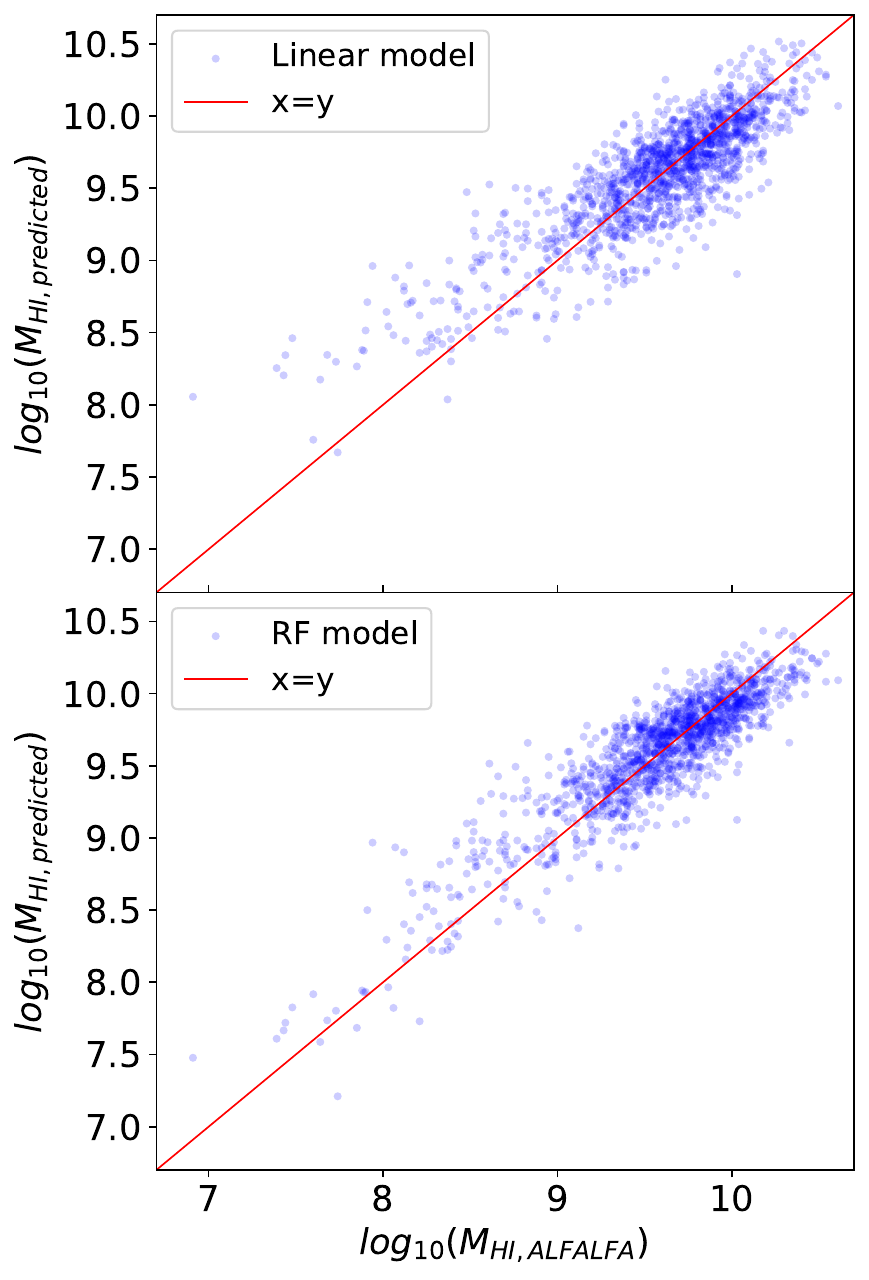}
	\caption{The scatter plot of the predicted and ALFALFA \Mhi evaluated on the test set for the linear model (upper panel) and the RF model (lower panel).}
	\label{RF_vs_lin_scatterplot}
\end{figure}

\begin{figure}
	\centering
	\includegraphics[width=0.45\textwidth]{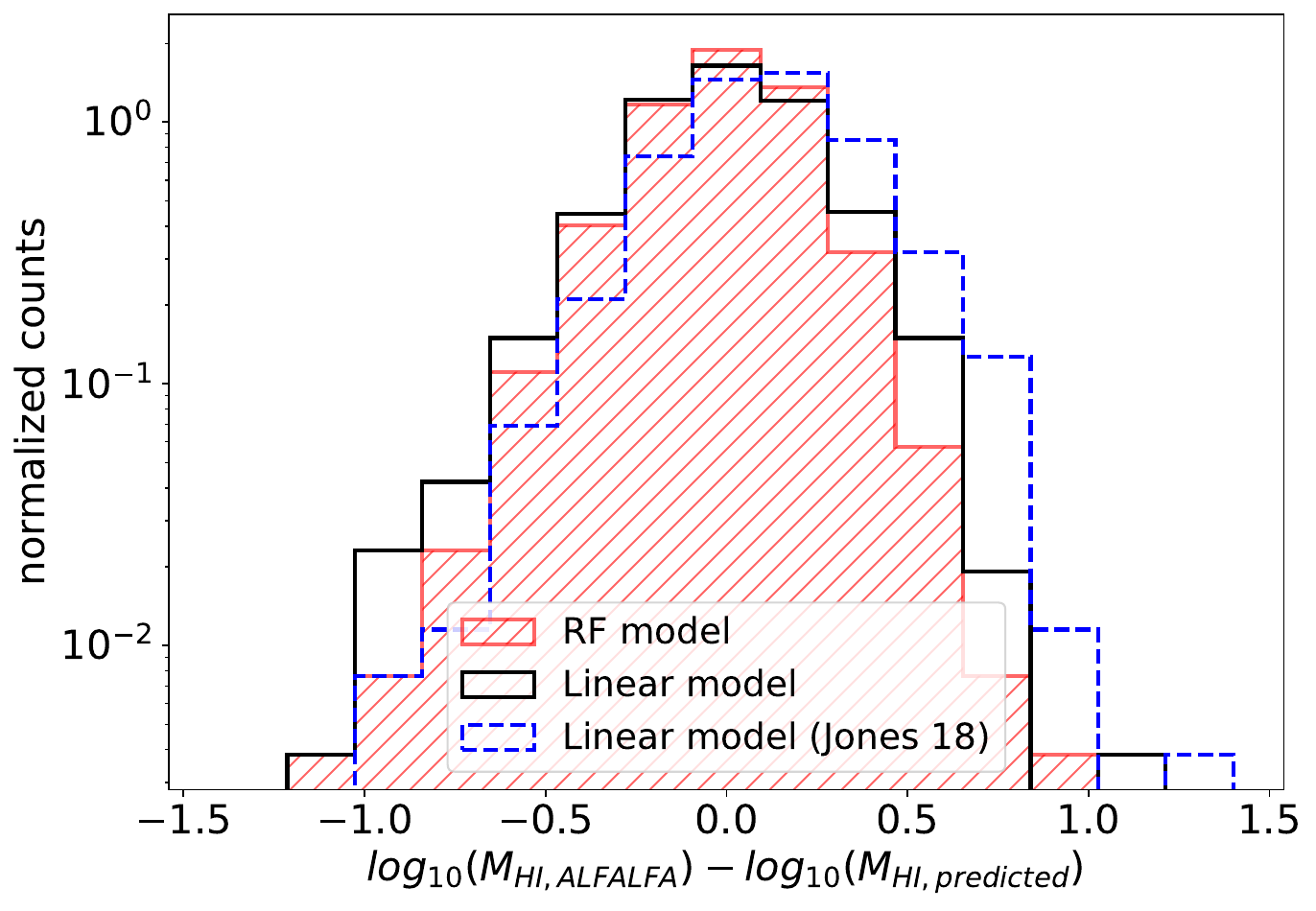}
	\caption{The distributions of residuals ($\mathrm{log}_{10}M_{HI,ALFALFA} - \mathrm{log}_{10}M_{HI,predicted}$) evaluated on the test set for the RF model (red, hatched histogram), linear model (black line histogram) and \cite{2018A_A...609A..17J} linear model (blue, dashed line histogram).}
	\label{RF_vs_lin_res_hist}
\end{figure}

We illustrate the performance on the test set for the linear and RF model in Figs. \ref{RF_vs_lin_scatterplot} and \ref{RF_vs_lin_res_hist}. Fig.\ref{RF_vs_lin_scatterplot} compares the predicted and ALFALFA \Mhi for both models, showing that the predictions of the RF model are clustered closer to the x=y line than the linear model predictions. Fig.\ref{RF_vs_lin_res_hist} presents the distributions of residuals for both models, defined as the difference between the observed ALFALFA \Mhi and the predicted \Mhi (we also plot the residuals obtained from \cite{2018A_A...609A..17J} linear model for comparison). The distribution of residuals for the RF predictions is narrower than the distributions for linear models. This reduction in scatter is the main advantage of the RF method: by capturing nonlinear dependencies between optical properties and \Mhi, it provides a more precise estimate of the expected \hi mass.

The relative importances of model features are listed in Table \ref{feat_imp}. As discussed above, the impurity based and permutation feature importances are moderately different. The strongest predictors are absolute magnitudes in g and r bands and the Petrosian radius $R_{90,g}$. Overall, Petrosian radii containing 90\% of the flux in all bands are significantly more important for the model than physical Petrosian radii or the Petrosian radii containing 50\% of the flux. Generally, relative feature importances of the model are in fine agreement with the correlation analysis discussed in Sect. \ref{ch3.2}.

\begin{table}
	\centering
	\caption{Permutation and impurity based relative feature importances of our RF model (only $ >2\% $ are listed).}
	\begin{tabular}{cc|cc}
		\hline
		\multicolumn{2}{c|}{Impurity based} & \multicolumn{2}{c}{Permutation based} \\
		\hline\hline
		g & 38.9\% & g & 41.9\%  \\
		\hline
		r & 18.0\% & $R_{90,g}$ & 31.3\%  \\
		\hline
		$R_{90,g}$ & 14.7\% & r & 11.1\%  \\
		\hline
		$R_{90,i}$ & 7.7\% & $R_{90,r}$ & 4.1\%  \\
		\hline
		$R_{90,r}$ & 3.8\% & r - i color & 3.4\%  \\
		\hline
		r - i color & 2.3\% & $R_{90,i}$ & 3.3\%  \\
		\hline
		i & 2.1\% & & \\
		\hline
	\end{tabular}\label{key}
	\label{feat_imp}
\end{table}

\section{\hi deficiency}
\label{ch4}

In this Section we use the RF model to predict the expected \hi mass and calculate the \hi deficiency (Eq. \ref{HIdef}) for non-isolated galaxies (nIG) from the ALFALFA sample. We also examine the relations between \hi deficiency and different tracers of the galactic environment. Throughout, the goal is not to perform a detailed environmental analysis, but rather to verify that our predictions of the expected (unaltered) \Mhi behave consistently when applied to galaxies in a wide range of environments. This serves as a demonstration that our method reproduces the broad trends established in previous studies.

Besides environmental effects the \hi content of a galaxy can be modified by internal factors. As we aim to study effects of the galactic environment, we mitigated the influence of internal factors by removing galaxies with AGN from our analysis and splitting the dataset into low-mass and high-mass galaxies, based on the stellar mass. Galaxies with $\mathrm{log}_{10}\frac{M_*}{M_\odot}<10.5$ act qualitatively different than higher stellar mass galaxies as the gas content varies with the stellar mass \citep{2017ApJS..233...22S}.

We used the RF model described in Sect. \ref{ch3.4} trained on the full IG sample (6\,982 objects, see Sect. \ref{ch3.2}) to predict the expected \Mhi for nIG. The nIG sample (8\,232 objects) was constructed from ALFALFA-SDSS sources which have available environmental information (see Sect. \ref{ch2.3}), excluding IG ($N_{gal}=1$) and galaxies with AGN. We also removed all galaxies with feature values outside the range of the IG training set (25 galaxies in total) as the RF model cannot extrapolate. A comparison between the IG and nIG samples is provided in the Appendix \ref{Appendix_3}. The range of applicability of our model, as sampled by the training set (IG sample), is given in Table \ref{samp_tab} for each feature. We provide the expected (predicted) \Mhi for nIG sample in a catalog available at Zenodo \citep{catalog}. It consists of following columns:
\begin{enumerate}
	\item \textbf{ALFALFA\_ID}: entry number in the Arecibo General Catalog (AGC);
	\item \textbf{SDSS\_ID}: SDSS DR15 photometric object identification number;
	\item \textbf{logMHI\_o}: observed logarithm of the \hi mass in solar units from ALFALFA catalog;
	\item \textbf{logMHI\_e}: expected logarithm of the \hi mass in solar units, predicted using RF model.
\end{enumerate}
Finally, the resulting nIG sample was split into galaxies with stellar mass $\mathrm{log}_{10}\frac{M_*}{M_\odot}<10.5$ (6\,991 objects) and $\mathrm{log}_{10}\frac{M_*}{M_\odot} \geq 10.5$ (1\,241 objects). 

\begin{figure}
	\centering
	\includegraphics[width=0.5\textwidth]{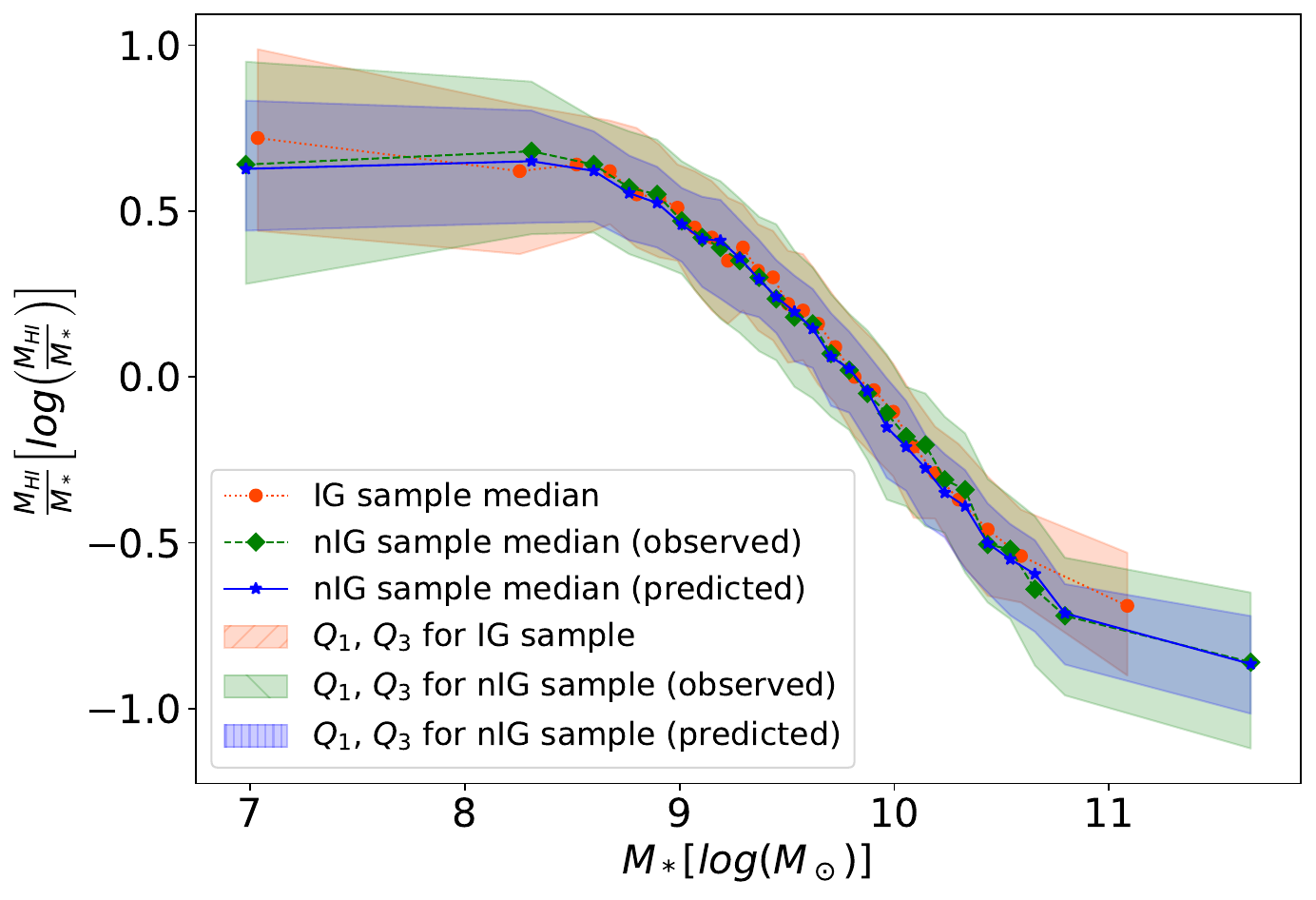}
	\caption{The distributions of the \hi to stellar mass ratio for the IG sample (red) and the nIG sample with observed (green) and predicted (blue) \Mhi, using equally populated bins. The shaded area represents the 1st and 3rd quartiles of distributions.}
	\label{dist_MHI_Ms}
\end{figure}

Before analyzing any environmental trends, we first explore the validity of predicted \Mhi. To verify that our predictions of expected \hi mass remain reliable across the entire range of stellar masses, we compare the distributions of the \hi to stellar mass ratio for the IG sample (the training set) and the nIG sample, considering both the observed and predicted \Mhi (Fig. \ref{dist_MHI_Ms}). The IG distribution closely aligns with both the observed and predicted nIG distributions, indicating that the training sample is appropriate for deriving the predictive model and confirming the overall consistency and robustness of our approach.

\begin{figure}
	\centering
	\includegraphics[width=0.45\textwidth]{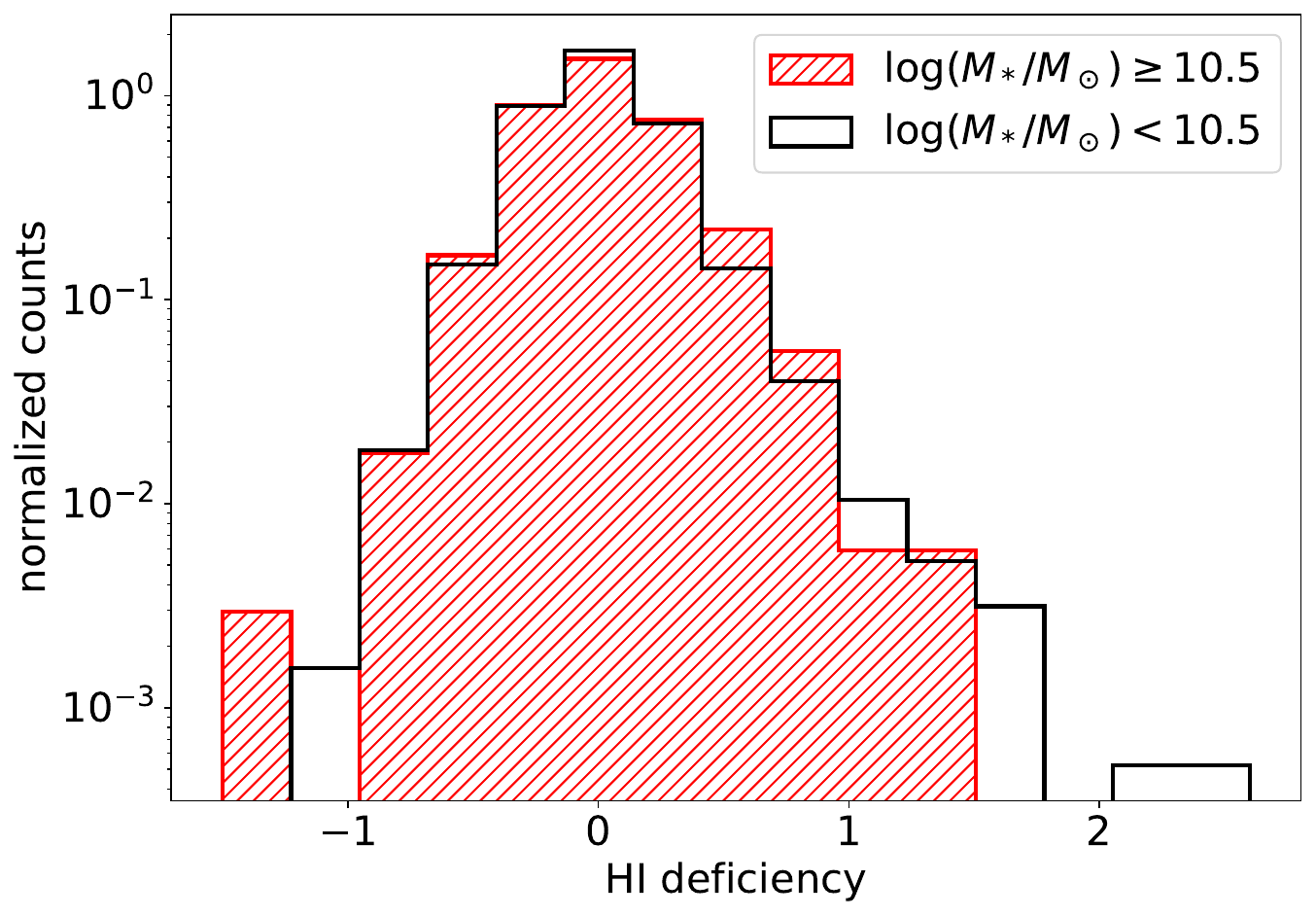}
	\caption{\hi deficiency histograms for the low-mass and high-mass nIG subsamples.}
	\label{def_hist}
\end{figure}

In Fig. \ref{def_hist}, we present the \hi deficiency histograms for the low-mass and high-mass subsamples. The two distributions span a very similar range of deficiency values with almost the same profiles, indicating that our results are not strongly limited by stellar mass. In particular, this suggests that \hi mass selection effects should not significantly bias our sample. 

\begin{figure*}
	\centering
	\includegraphics[width=0.85\textwidth]{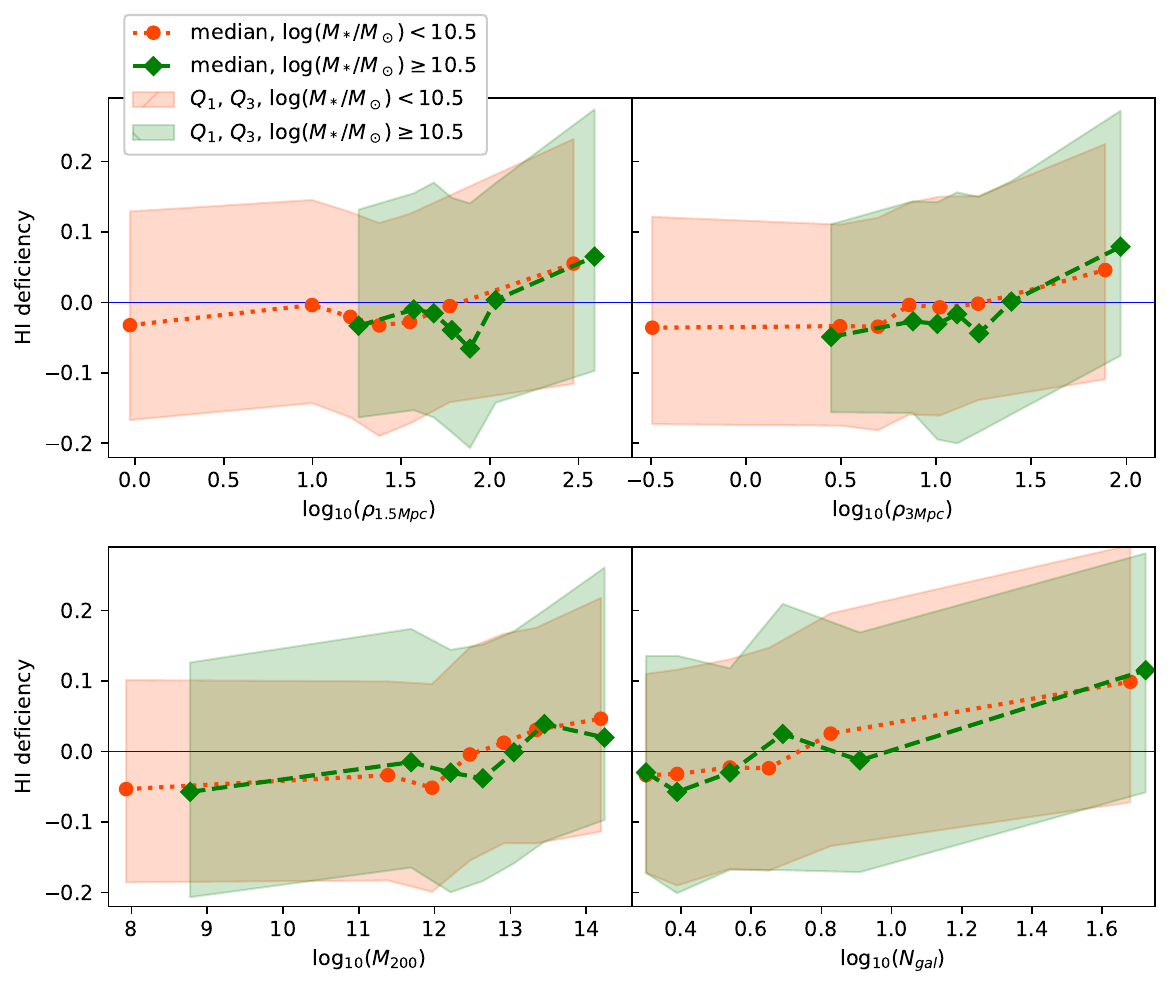}
	\caption{The binned distribution of the \hi deficiency with respect to different environmental tracers; orange circles represent median values for low-mass galaxies ($\mathrm{log}_{10}\frac{M_*}{M_\odot}<10.5$) and green squares denote median values for high-mass galaxies ($\mathrm{log}_{10}\frac{M_*}{M_\odot}\geq10.5$). The shaded area represents the 1st and 3rd quartiles of distributions. Galaxies in each subsample are binned into seven equally populated bins.}
	\label{def_err}
\end{figure*}

Fig. \ref{def_err} shows a calculated median of the \hi deficiency for the subsample of low-mass (orange) and high-mass galaxies (green) with respect to the four environmental tracers (as introduced in Sect. \ref{ch2.3}). Galaxies in each subsample are binned into seven equally populated bins. The x-axis position of each bin is determined as the middle value between the bin edges. The shaded area represents the 1st and 3rd quartiles for each distribution. Since bins for given subsample contain equal numbers of galaxies, they all have the same statistical significance. 

The binned distributions in Fig. \ref{def_err} show that the \hi deficiency is related with the galactic environment, with the median value increasing from around -0.05 at sparse environments up to 0.1 at the densest environments. The scatter is large (difference between 1st and 3rd quartile around 0.3), reflecting measurement errors in galaxy properties and environmental tracers as well as uncertainties in model predictions (for further discussion about the scatter, see Sect. \ref{ch5}). These sources of noise broaden the distributions but should not systematically shift them. Although the \hi deficiency with respect to various tracers expresses moderately different behaviours, there is a prevailing growing trend. This indicates, that the environment depletes the neutral hydrogen gas from galaxies. 

Overall, the behavior of \hi deficiency with environmental tracers is consistent with established trends from previous studies. This agreement indicates that our method for estimating the expected (unaltered) \hi mass produces physically plausible results, supporting the reliability of our model for further applications.

\subsection{Time evolution of \hi deficiency} \label{time_evolution}

\begin{figure*}
	\centering
	\includegraphics[width=1\textwidth]{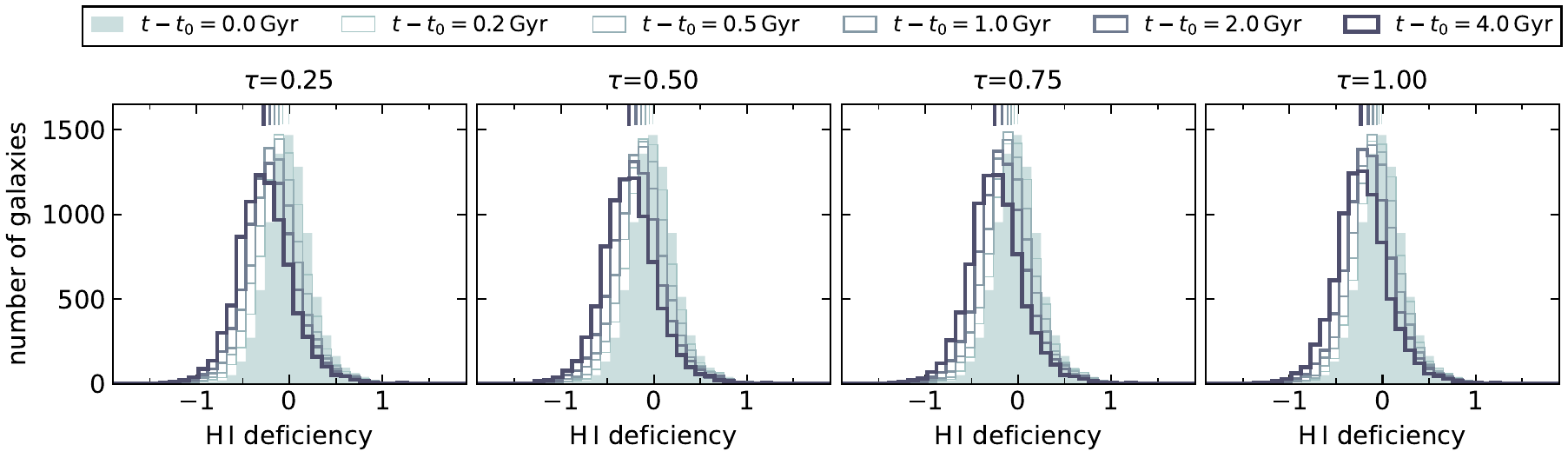}
	\caption{The distribution of \hi deficiencies (computed with the RF model) for the nIG sample (solid histograms) and their expected evolution driven by the aging stellar populations (empty histograms). Separate panels show the expected evolution result of exponentially declining star formation rate with $\tau$ indicated in the panel titles. The medians of the distributions are marked by the small vertical bars at the top.}
	\label{def_evolution}
\end{figure*}

The mapping of \hi mass onto optical properties done by the RF model in Sect. \ref{ch3.4} uses all 17 measured optical parameters. The relative importances of the 17 most important parameters are listed in Table \ref{feat_imp}. The optical properties of a galaxy are determined by its star formation which is indirectly determined by its \hi content. A galaxy with a steady state star formation rate will have its optical properties in equilibrium with its \hi content. The optical properties will follow any changes in the gas content with a delay, driven by the timescales on which the stellar populations evolve. Thus any gas removal event (particularly rapid one) will necessarily break this equilibrium, as the optical emission will still be dominated by the presence of young stars. As the stellar population ages a new equilibrium will be established, valid for galaxies without detectable \hi, which are not present in our data set. In essence, when using this model to estimate \hi deficiency we are potentially selecting galaxies whose optical properties are not in equilibrium with their \hi content (see \citealt{Taylor2020} for a related discussion). This implies that the estimated \hi deficiency will evolve with time, as the optical parameters on which it is based evolve.

As shown on Fig. \ref{corr_matrice} most of the input parameters to our model correlate not only with the \hi mass of the galaxy but also with many of the other input parameters. This makes a direct calculation of the way the predicted \hi mass will evolve, as the stellar population of the galaxy evolves, difficult. Because of that we decided to simulate the evolution of the stellar population of galaxies with different star formation histories and see how this evolution will impact on their predicted \hi  mass. Below, we present a ``toy model" aimed at providing a qualitative, conceptual demonstration aiming to give a rough idea of the magnitude of the effect, not a rigorous prediction.

For the simulation we use the “Differentiable Stellar Population Synthesis” (DSPS, \citealt{Hearin_2023}) package for python. The setup is a constant star formation rate for the first 5 Gyrs, followed by an exponential decline in the star formation rate starting at $t_{0}$ as $\mathrm{SFR} = \exp[-(t - t_0)/\tau]$, with $\tau \in \{0.25,0.5,0.75,1.0\}\,\mathrm{Gyr}$. The separate star formation histories and optical parameters derived from DSPS are shown on Fig. E.1, see appendix \ref{Appendix_4}. This choice of $\tau$ values samples different physical gas-removal mechanisms. At the low end, strong ram-pressure stripping which is expected to remove all the gas and lead to complete stoppage of star formation on time-scales of $< 1$ Gyr. At the high end, $\tau = 1.0$ samples the expected decline in star formation resulting from stoppage of gas accretion (starvation, \citealt{Larson1980}), reaching complete stoppage of star formation over 4 Gyrs \citep{2015Natur.521..192P}. DSPS gives the evolution of rest-frame g/r/i magnitudes which we also use to calculate the expected color evolution. 

Another very important optical parameter used by the RF model is the optical radius. Although a significant amount of research has been done on the evolution of the sizes of passive and star forming galaxies at different environments \citep{Kuchner_2017}, at different redshifts \citep{Kelvin_2012} and in simulations \citep{Genel_2018} we could not find a prescription that will allow us to estimate the size evolution of a galaxy, measured in g/r/i filters, following different gas-stripping scenarios. Because of that we present on Fig. \ref{def_evolution} only the effects that evolving luminosity has on the predicted \hi mass noting that this is likely a lower limit to the actual effect. Given that gas removal is likely to induce outside-in quenching of the star formation, it is likely that there will be substantial reduction in galaxy size measured in the bluer bands, which will further decrease the predicted \hi mass.

Figure \ref{def_evolution} shows the distribution of \hi deficiencies (computed with the RF model) for the nIG sample (filled histograms) and the expected evolution over 4 Gyrs, following star formation quenching on different time-scales (columns). Over the first 1 Gyr since the stoppage of star formation has began, the decrease in predicted \hi deficiency is between 0.1 for the slowest and 0.16 for the fastest declining star formation rate. After 4 Gyrs the predicted \hi deficiency will drop by 0.23 and 0.28. This is enough to reduce the number of galaxies with positive \hi deficiency in the nIG sample by $\sim$ 45\% over the first 1 Gyr and $\sim$ 66\% over 4 Gyrs, assuming $\tau=0.25$ and that they are all observed at the onset of gas stripping. The main conclusion from this experiment is that \hi deficiency is only discernible for a certain period of time. This effect could also contribute some of the observed scatter in expected \Mhi for IGs.

We also explored the effects of a range of metallicities on the evolution of the optical luminosity but found it did not alter the results significantly. The results shown in Fig.\ref{def_evolution} are obtained with solar metallicity. While not exhaustive, this analysis shows that the estimated \hi deficiency will depend on the time since the gas removal began and on how rapid the gas removal is, as well as the nature of the gas removal mechanism. Note that these arguments will likely be true for both linear models of \cite{1984AJ.....89..758H} and \cite{2018A_A...609A..17J} as they map \hi mass to fixed surface-brightness B-band optical size, a parameter which will also likely evolve as the stellar population evolves. 

\section{Discussion and conclusion}
\label{ch5}

The \hi content of isolated galaxies is related to their optical properties. From 17 inspected optical quantities, the absolute magnitude in g band shows the strongest correlation with \Mhi and is the most important predictor of \Mhi in our RF model. This is expected, since the g band light is dominated by young stars which track the star formation rate that depends on \Mhi. The morphology type does not correlate with the \hi mass in our IG sample. However, this is rather caused by observational limitations than by intrinsic astrophysical mechanisms, as this work is based on \hi selected galaxies which have relatively constant concentration indices and broad-band colors.

Although being a good first approximation, the linear relation between \Mhi and the optical diameter cannot match the performance of the RF model. With $ \mathrm{RMSE} \approx 0.22~\mathrm{dex} $ and $ R^2 \approx 0.80 $, our RF model noticeably outperforms the linear method which scored $ \mathrm{RMSE} \approx 0.26~\mathrm{dex} $ and $ R^2 \approx 0.70 $. A similar conclusion was reached by \cite{2017MNRAS.464.3796T}, who also reported a reduction in scatter when moving from linear to machine learning models (from $\sim$0.32 to $\sim$0.22 dex). Although their study focused on predicting the gas mass fraction and employed artificial neural networks with a different set of input features (they did not restrict the sample to isolated galaxies; instead including environmental tracer as a predictor), the overall trend is consistent with our findings. \cite{2020ApJ...900..142W} predicted the gas mass fraction from SDSS $g$, $r$, and $i$-band images using convolutional neural networks, achieving an RMSE of 0.23 dex. While random forests are not well suited for image-based inputs, they also reported a comparison with an RF model, which resulted in a higher RMSE of 0.31 dex.

The main advantage of our work lies in the use of a highly homogeneous dataset combined with a sample size that is several times larger than those used in previous studies of \hi deficiency. While \cite{2018A_A...609A..17J} achieved a significant reduction in intrinsic scatter through extremely strict isolation criteria and an advanced regression analysis that incorporated upper limits, their approach necessarily relied on a much smaller control sample. Our method is complementary: although our isolation criterion is less strict, the homogeneity of the data and the substantially larger sample enable a more statistically robust characterization of the expected \hi content, yielding a more broadly applicable predictive framework.

We find an increase in binned median \hi deficiency of 0.15 dex attributable to environmental effects. We note that the scatter in \hi deficiency with respect to the environment remains substantial (difference between 1st and 3rd quartile around 0.3 dex). It is consistent with previous studies, although our method provides an improvement over earlier approaches. This scatter reflects not only measurement errors and model uncertainties but also the long timescales over which environmental mechanisms act. Processes such as RPS, starvation, or tidal interactions operate over long periods - typically on the order of at least a few $10^8$ years for RPS and up to several $10^9$ years for starvation \citep{2006PASP..118..517B,2015Natur.521..192P}. Since galaxies move through different regions, often toward denser environments, populations at various stages of gas depletion and morphological transformation can coexist within the same environment. Consequently, it is possible to find galaxies that still contain \hi and show ongoing star formation even in high-density regions \citep{2022A&A...657A...9C}. These galaxies have likely entered their environment only recently, within a timescale shorter than that required for significant gas removal. This gradual transformation process likely contributes to a substantial portion of the observed scatter in the \hi deficiency.

Despite considerable scatter, the increasing trend of the \hi deficiency with respect to the environment is recognizable. Both the scatter and the overall trend of the \hi deficiency are qualitatively consistent with results reported by \cite{2008MNRAS.383.1519C}. Furthermore, the small difference between the low-mass and high-mass curves indicates that the bulk of the deficiency evolution is driven by environment, rather than secular evolution with stellar mass, in agreement with the findings of \cite{2009MNRAS.393.1324B}.

As noted throughout this paper, our work has several limitations. Since we used the environment catalog based on spectroscopy, our sample is biased towards brighter galaxies. With a sky coverage of approximately 7\,000 square degrees, ALFALFA provides an excellent sampling of a wide range of environments, so no significant bias related to sky area is expected. Like all flux-limited surveys, ALFALFA is affected by the Malmquist bias, which favors intrinsically brighter (more \hi rich) galaxies at larger distances. In principle, this bias could be mitigated by imposing a distance cut to restrict the analysis to a volume where the survey is approximately mass-complete. However, we chose not to apply such a cut in order to retain a larger sample. Moreover, because ALFALFA spans a statistically large but cosmologically small volume, we do not expect any significant negative impact of the Malmquist bias: low-mass galaxies are detected with high completeness only at the nearest distances, whereas high-mass galaxies reach high completeness at larger distances. Since no redshift evolution is expected over the limited range probed by ALFALFA, galaxies of all masses are detected at high completeness levels somewhere within the survey volume, without affecting their intrinsic physical properties. Finally, the distribution of our sample is heavily shifted towards LTGs. Although those biases limit the applicability of our model to only gas-rich galaxies, the same that are present in the training set, they do not affect the predictions made for them. The RF algorithm has two intrinsic limitations: it does not account for the measurement errors and it cannot extrapolate. Therefore we applied our model only to galaxies which have all features lying within the range sampled by the galaxies present in the training set.

Future work will focus on extending both the dataset and the modeling framework. In particular, incorporating larger \hi samples from forthcoming surveys will improve the statistical robustness and environmental coverage of the analysis. We also plan to expand the feature space by including additional optical parameters, exploiting deeper imaging data, and refining structural measurements such as galaxy radii. Further efforts will aim to better control selection effects and reduce biases in the sample. Finally, we will address the current lack of uncertainty treatment in ML models by exploring probabilistic extensions of the random forest algorithm, which allow measurement errors in both input features and target variables to be naturally incorporated, potentially leading to more robust predictions.

\section*{Acknowledgments}

FJ was supported by the Visegrad Fund grant No. 22210105 and by the VEGA – the Slovak Grant Agency for Science, grant No. 1/0481/25. BD acknowledges funding from the HTM (grant TK202) and the EU Horizon Europe (EXCOSM, grant No. 101159513), as well as from the Estonian Research Council (grant PSG1200). RN was funded by the EU NextGenerationEU through the Recovery and Resilience Plan for Slovakia under the project No. 09I03-03-V04-00137. RT was supported by the institutional project RVO:67985815 and the Czech Ministry of Education, Youth and Sports from the large infrastructures for Research, Experimental Development and Innovations project LM 2015067.

ALFALFA observations were obtained with the Arecibo Observatory. At the time the data were taken, the Arecibo Observatory was part of the National Astronomy and Ionosphere Center (NAIC), operated by Cornell University under a cooperative agreement with the National Science Foundation (NSF). These data products are the work of the entire ALFALFA collaboration who have contributed to the many aspects of the survey over the years. Their research was supported by grants from a variety of agencies, including the NSF and the Brinson Foundation. 

Funding for the Sloan Digital Sky Survey IV has been provided by the Alfred P. Sloan Foundation, the U.S. Department of Energy Office of Science, and the Participating Institutions. SDSS acknowledges support and resources from the Center for High-Performance Computing at the University of Utah. The SDSS web site is www.sdss4.org.SDSS is managed by the Astrophysical Research Consortium for the Participating Institutions of the SDSS Collaboration including the Brazilian Participation Group, the Carnegie Institution for Science, Carnegie Mellon University, Center for Astrophysics | Harvard \& Smithsonian (CfA), the Chilean Participation Group, the French Participation Group, Instituto de Astrofísica de Canarias, The Johns Hopkins University, Kavli Institute for the Physics and Mathematics of the Universe (IPMU) / University of Tokyo, the Korean Participation Group, Lawrence Berkeley National Laboratory, Leibniz Institut für Astrophysik Potsdam (AIP), Max-Planck-Institut für Astronomie (MPIA Heidelberg), Max-Planck-Institut für Astrophysik (MPA Garching), Max-Planck-Institut für Extraterrestrische Physik (MPE), National Astronomical Observatories of China, New Mexico State University, New York University, University of Notre Dame, Observatório Nacional / MCTI, The Ohio State University, Pennsylvania State University, Shanghai Astronomical Observatory, United Kingdom Participation Group, Universidad Nacional Autónoma de México, University of Arizona, University of Colorado Boulder, University of Oxford, University of Portsmouth, University of Utah, University of Virginia, University of Washington, University of Wisconsin, Vanderbilt University, and Yale University.

\bibliographystyle{aasjournalv7}
\bibliography{ref}

@ARTICLE{2018ApJ...861...49H,
       author = {{Haynes}, Martha P. and {Giovanelli}, Riccardo and {Kent}, Brian R. and {Adams}, Elizabeth A.~K. and {Balonek}, Thomas J. and {Craig}, David W. and {Fertig}, Derek and {Finn}, Rose and {Giovanardi}, Carlo and {Hallenbeck}, Gregory and {Hess}, Kelley M. and {Hoffman}, G. Lyle and {Huang}, Shan and {Jones}, Michael G. and {Koopmann}, Rebecca A. and {Kornreich}, David A. and {Leisman}, Lukas and {Miller}, Jeffrey and {Moorman}, Crystal and {O'Connor}, Jessica and {O'Donoghue}, Aileen and {Papastergis}, Emmanouil and {Troischt}, Parker and {Stark}, David and {Xiao}, Li},
        title = "{The Arecibo Legacy Fast ALFA Survey: The ALFALFA Extragalactic H I Source Catalog}",
      journal = {\apj},
         year = 2018,
        month = jul,
       volume = {861},
       number = {1},
          eid = {49},
        pages = {49},
          doi = {10.3847/1538-4357/aac956},
archivePrefix = {arXiv},
       eprint = {1805.11499},
 primaryClass = {astro-ph.GA},
       adsurl = {https://ui.adsabs.harvard.edu/abs/2018ApJ...861...49H}
}

@ARTICLE{1984AJ.....89..758H,
	author = {{Haynes}, M.~P. and {Giovanelli}, R.},
	title = "{Neutral hydrogen in isolated galaxies. IV. Results for the Arecibo sample.}",
	journal = {\aj},
	year = 1984,
	month = jun,
	volume = {89},
	pages = {758-800},
	doi = {10.1086/113573},
	adsurl = {https://ui.adsabs.harvard.edu/abs/1984AJ.....89..758H}
}

@ARTICLE{2022A&A...665A.155D,
	author = {{Deshev}, Boris and {Taylor}, Rhys and {Minchin}, Robert and {Scott}, Tom C. and {Brinks}, Elias},
	title = "{The Arecibo Galaxy Environment Survey (AGES). XI. The expanded Abell 1367 field: Data catalogue and H I census over the surveyed volume}",
	journal = {\aap},
	year = 2022,
	month = sep,
	volume = {665},
	eid = {A155},
	pages = {A155},
	doi = {10.1051/0004-6361/202243103},
	archivePrefix = {arXiv},
	eprint = {2206.13533},
	primaryClass = {astro-ph.GA},
	adsurl = {https://ui.adsabs.harvard.edu/abs/2022A&A...665A.155D}
}

@ARTICLE{1983AJ.....88..881G,
	author = {{Giovanelli}, R. and {Haynes}, M.~P.},
	title = "{The HI extent and deficiency of spiral galaxies in the Virgo cluster.}",
	journal = {\aj},
	year = 1983,
	month = jul,
	volume = {88},
	pages = {881-908},
	doi = {10.1086/113376},
	adsurl = {https://ui.adsabs.harvard.edu/abs/1983AJ.....88..881G}
}

@ARTICLE{2010A&A...518L..49C,
	author = {{Cortese}, L. and {Davies}, J.~I. and {Pohlen}, M. and {Baes}, M. and {Bendo}, G.~J. and {Bianchi}, S. and {Boselli}, A. and {De Looze}, I. and {Fritz}, J. and {Verstappen}, J. and {Bomans}, D.~J. and {Clemens}, M. and {Corbelli}, E. and {Dariush}, A. and {di Serego Alighieri}, S. and {Fadda}, D. and {Garcia-Appadoo}, D.~A. and {Gavazzi}, G. and {Giovanardi}, C. and {Grossi}, M. and {Hughes}, T.~M. and {Hunt}, L.~K. and {Jones}, A.~P. and {Madden}, S. and {Pierini}, D. and {Sabatini}, S. and {Smith}, M.~W.~L. and {Vlahakis}, C. and {Xilouris}, E.~M. and {Zibetti}, S.},
	title = "{The Herschel Virgo Cluster Survey . II. Truncated dust disks in H I-deficient spirals}",
	journal = {\aap},
	year = 2010,
	month = jul,
	volume = {518},
	eid = {L49},
	pages = {L49},
	doi = {10.1051/0004-6361/201014550},
	archivePrefix = {arXiv},
	eprint = {1005.3055},
	primaryClass = {astro-ph.CO},
	adsurl = {https://ui.adsabs.harvard.edu/abs/2010A&A...518L..49C}
}

@ARTICLE{2009MNRAS.400.1962K,
	author = {{Kilborn}, Virginia A. and {Forbes}, Duncan A. and {Barnes}, David G. and {Koribalski}, B{\"a}rbel S. and {Brough}, Sarah and {Kern}, Katie},
	title = "{Southern GEMS groups - II. HI distribution, mass functions and HI deficient galaxies}",
	journal = {\mnras},
	year = 2009,
	month = dec,
	volume = {400},
	number = {4},
	pages = {1962-1985},
	doi = {10.1111/j.1365-2966.2009.15587.x},
	archivePrefix = {arXiv},
	eprint = {0909.0568},
	primaryClass = {astro-ph.CO},
	adsurl = {https://ui.adsabs.harvard.edu/abs/2009MNRAS.400.1962K}
}

@ARTICLE{1991A&A...249..359C,
	author = {{Casoli}, F. and {Boisse}, P. and {Combes}, F. and {Dupraz}, C.},
	title = "{Are HI-deficient galaxies of the Coma supercluster deficient in moleculargas ?}",
	journal = {\aap},
	year = 1991,
	month = sep,
	volume = {249},
	pages = {359},
	adsurl = {https://ui.adsabs.harvard.edu/abs/1991A&A...249..359C}
}

@ARTICLE{1997A&A...325..473H,
	author = {{Huchtmeier}, W.~K.},
	title = "{HI-deficiency in Hickson compact groups of galaxies.}",
	journal = {\aap},
	year = 1997,
	month = sep,
	volume = {325},
	pages = {473-478},
	adsurl = {https://ui.adsabs.harvard.edu/abs/1997A&A...325..473H}
}

@ARTICLE{1985A&A...151..108G,
	author = {{Guiderdoni}, B. and {Rocca-Volmerange}, B.},
	title = "{Evolution of spiral galaxies in the Virgo cluster. I. Statistical analysis of HI deficiency and colors.}",
	journal = {\aap},
	year = 1985,
	month = oct,
	volume = {151},
	pages = {108-120},
	adsurl = {https://ui.adsabs.harvard.edu/abs/1985A&A...151..108G}
}

@ARTICLE{2006PASP..118..517B,
	author = {{Boselli}, Alessandro and {Gavazzi}, Giuseppe},
	title = "{Environmental Effects on Late-Type Galaxies in Nearby Clusters}",
	journal = {\pasp},
	year = 2006,
	month = apr,
	volume = {118},
	number = {842},
	pages = {517-559},
	doi = {10.1086/500691},
	archivePrefix = {arXiv},
	eprint = {astro-ph/0601108},
	primaryClass = {astro-ph},
	adsurl = {https://ui.adsabs.harvard.edu/abs/2006PASP..118..517B}
}

@ARTICLE{2020AJ....160..271D,
	author = {{Durbala}, Adriana and {Finn}, Rose A. and {Crone Odekon}, Mary and {Haynes}, Martha P. and {Koopmann}, Rebecca A. and {O'Donoghue}, Aileen A.},
	title = "{The ALFALFA-SDSS Galaxy Catalog}",
	journal = {\aj},
	year = 2020,
	month = dec,
	volume = {160},
	number = {6},
	eid = {271},
	pages = {271},
	doi = {10.3847/1538-3881/abc018},
	archivePrefix = {arXiv},
	eprint = {2011.02588},
	primaryClass = {astro-ph.GA},
	adsurl = {https://ui.adsabs.harvard.edu/abs/2020AJ....160..271D}
}

@ARTICLE{2016A&A...588A..14T,
	author = {{Tempel}, E. and {Kipper}, R. and {Tamm}, A. and {Gramann}, M. and {Einasto}, M. and {Sepp}, T. and {Tuvikene}, T.},
	title = "{Friends-of-friends galaxy group finder with membership refinement. Application to the local Universe}",
	journal = {\aap},
	year = 2016,
	month = apr,
	volume = {588},
	eid = {A14},
	pages = {A14},
	doi = {10.1051/0004-6361/201527755},
	archivePrefix = {arXiv},
	eprint = {1601.01117},
	primaryClass = {astro-ph.CO},
	adsurl = {https://ui.adsabs.harvard.edu/abs/2016A&A...588A..14T}
}

@ARTICLE{2018A&A...618A..81T,
	author = {{Tempel}, Elmo and {Kruuse}, Maarja and {Kipper}, Rain and {Tuvikene}, Taavi and {Sorce}, Jenny G. and {Stoica}, Radu S.},
	title = "{Bayesian group finder based on marked point processes. Method and feasibility study using the 2MRS data set}",
	journal = {\aap},
	year = 2018,
	month = oct,
	volume = {618},
	eid = {A81},
	pages = {A81},
	doi = {10.1051/0004-6361/201833217},
	archivePrefix = {arXiv},
	eprint = {1806.04469},
	primaryClass = {astro-ph.CO},
	adsurl = {https://ui.adsabs.harvard.edu/abs/2018A&A...618A..81T}
}

@ARTICLE{2021ApJ...909..143Y,
	author = {{Yang}, Xiaohu and {Xu}, Haojie and {He}, Min and {Gu}, Yizhou and {Katsianis}, Antonios and {Meng}, Jiacheng and {Shi}, Feng and {Zou}, Hu and {Zhang}, Youcai and {Liu}, Chengze and {Wang}, Zhaoyu and {Dong}, Fuyu and {Lu}, Yi and {Li}, Qingyang and {Chen}, Yangyao and {Wang}, Huiyuan and {Mo}, Houjun and {Fu}, Jian and {Guo}, Hong and {Leauthaud}, Alexie and {Luo}, Yu and {Zhang}, Jun and {Zu}, Ying},
	title = "{An Extended Halo-based Group/Cluster Finder: Application to the DESI Legacy Imaging Surveys DR8}",
	journal = {\apj},
	year = 2021,
	month = mar,
	volume = {909},
	number = {2},
	eid = {143},
	pages = {143},
	doi = {10.3847/1538-4357/abddb2},
	archivePrefix = {arXiv},
	eprint = {2012.14998},
	primaryClass = {astro-ph.GA},
	adsurl = {https://ui.adsabs.harvard.edu/abs/2021ApJ...909..143Y}
}

@ARTICLE{2017A&A...602A.100T,
	author = {{Tempel}, E. and {Tuvikene}, T. and {Kipper}, R. and {Libeskind}, N.~I.},
	title = "{Merging groups and clusters of galaxies from the SDSS data. The catalogue of groups and potentially merging systems}",
	journal = {\aap},
	year = 2017,
	month = jun,
	volume = {602},
	eid = {A100},
	pages = {A100},
	doi = {10.1051/0004-6361/201730499},
	archivePrefix = {arXiv},
	eprint = {1704.04477},
	primaryClass = {astro-ph.CO},
	adsurl = {https://ui.adsabs.harvard.edu/abs/2017A&A...602A.100T}
}

@ARTICLE{2012ApJS..199...34W,
	author = {{Wen}, Z.~L. and {Han}, J.~L. and {Liu}, F.~S.},
	title = "{A Catalog of 132,684 Clusters of Galaxies Identified from Sloan Digital Sky Survey III}",
	journal = {\apjs},
	year = 2012,
	month = apr,
	volume = {199},
	number = {2},
	eid = {34},
	pages = {34},
	doi = {10.1088/0067-0049/199/2/34},
	archivePrefix = {arXiv},
	eprint = {1202.6424},
	primaryClass = {astro-ph.CO},
	adsurl = {https://ui.adsabs.harvard.edu/abs/2012ApJS..199...34W}
}

@ARTICLE{2012A&A...540A.106T,
	author = {{Tempel}, E. and {Tago}, E. and {Liivam{\"a}gi}, L.~J.},
	title = "{Groups and clusters of galaxies in the SDSS DR8. Value-added catalogues}",
	journal = {\aap},
	year = 2012,
	month = apr,
	volume = {540},
	eid = {A106},
	pages = {A106},
	doi = {10.1051/0004-6361/201118687},
	archivePrefix = {arXiv},
	eprint = {1112.4648},
	primaryClass = {astro-ph.CO},
	adsurl = {https://ui.adsabs.harvard.edu/abs/2012A&A...540A.106T}
}

@ARTICLE{2014ApJ...785..104R,
	author = {{Rykoff}, E.~S. and {Rozo}, E. and {Busha}, M.~T. and {Cunha}, C.~E. and {Finoguenov}, A. and {Evrard}, A. and {Hao}, J. and {Koester}, B.~P. and {Leauthaud}, A. and {Nord}, B. and {Pierre}, M. and {Reddick}, R. and {Sadibekova}, T. and {Sheldon}, E.~S. and {Wechsler}, R.~H.},
	title = "{redMaPPer. I. Algorithm and SDSS DR8 Catalog}",
	journal = {\apj},
	year = 2014,
	month = apr,
	volume = {785},
	number = {2},
	eid = {104},
	pages = {104},
	doi = {10.1088/0004-637X/785/2/104},
	archivePrefix = {arXiv},
	eprint = {1303.3562},
	primaryClass = {astro-ph.CO},
	adsurl = {https://ui.adsabs.harvard.edu/abs/2014ApJ...785..104R}
}

@ARTICLE{2012MNRAS.422.1835S,
	author = {{Serra}, Paolo and {Oosterloo}, Tom and {Morganti}, Raffaella and {Alatalo}, Katherine and {Blitz}, Leo and {Bois}, Maxime and {Bournaud}, Fr{\'e}d{\'e}ric and {Bureau}, Martin and {Cappellari}, Michele and {Crocker}, Alison F. and {Davies}, Roger L. and {Davis}, Timothy A. and {de Zeeuw}, P.~T. and {Duc}, Pierre-Alain and {Emsellem}, Eric and {Khochfar}, Sadegh and {Krajnovi{\'c}}, Davor and {Kuntschner}, Harald and {Lablanche}, Pierre-Yves and {McDermid}, Richard M. and {Naab}, Thorsten and {Sarzi}, Marc and {Scott}, Nicholas and {Trager}, Scott C. and {Weijmans}, Anne-Marie and {Young}, Lisa M.},
	title = "{The ATLAS$^{3D}$ project - XIII. Mass and morphology of H I in early-type galaxies as a function of environment}",
	journal = {\mnras},
	year = 2012,
	month = may,
	volume = {422},
	number = {3},
	pages = {1835-1862},
	doi = {10.1111/j.1365-2966.2012.20219.x},
	archivePrefix = {arXiv},
	eprint = {1111.4241},
	primaryClass = {astro-ph.CO},
	adsurl = {https://ui.adsabs.harvard.edu/abs/2012MNRAS.422.1835S}
}

@ARTICLE{2001AJ....122.1861S,
	author = {{Strateva}, Iskra and {Ivezi{\'c}}, {\v{Z}}eljko and {Knapp}, Gillian R. and {Narayanan}, Vijay K. and {Strauss}, Michael A. and {Gunn}, James E. and {Lupton}, Robert H. and {Schlegel}, David and {Bahcall}, Neta A. and {Brinkmann}, Jon and {Brunner}, Robert J. and {Budav{\'a}ri}, Tam{\'a}s and {Csabai}, Istv{\'a}n and {Castander}, Francisco Javier and {Doi}, Mamoru and {Fukugita}, Masataka and {Gy{\H{o}}ry}, Zsuzsanna and {Hamabe}, Masaru and {Hennessy}, Greg and {Ichikawa}, Takashi and {Kunszt}, Peter Z. and {Lamb}, Don Q. and {McKay}, Timothy A. and {Okamura}, Sadanori and {Racusin}, Judith and {Sekiguchi}, Maki and {Schneider}, Donald P. and {Shimasaku}, Kazuhiro and {York}, Donald},
	title = "{Color Separation of Galaxy Types in the Sloan Digital Sky Survey Imaging Data}",
	journal = {\aj},
	year = 2001,
	month = oct,
	volume = {122},
	number = {4},
	pages = {1861-1874},
	doi = {10.1086/323301},
	archivePrefix = {arXiv},
	eprint = {astro-ph/0107201},
	primaryClass = {astro-ph},
	adsurl = {https://ui.adsabs.harvard.edu/abs/2001AJ....122.1861S}
}

@ARTICLE{2001AJ....122.1238S,
	author = {{Shimasaku}, Kazuhiro and {Fukugita}, Masataka and {Doi}, Mamoru and {Hamabe}, Masaru and {Ichikawa}, Takashi and {Okamura}, Sadanori and {Sekiguchi}, Maki and {Yasuda}, Naoki and {Brinkmann}, Jon and {Csabai}, Istv{\'a}n and {Ichikawa}, Shin-Ichi and {Ivezi{\'c}}, Zeljko and {Kunszt}, Peter Z. and {Schneider}, Donald P. and {Szokoly}, Gyula P. and {Watanabe}, Masaru and {York}, Donald G.},
	title = "{Statistical Properties of Bright Galaxies in the Sloan Digital Sky Survey Photometric System}",
	journal = {\aj},
	year = 2001,
	month = sep,
	volume = {122},
	number = {3},
	pages = {1238-1250},
	doi = {10.1086/322094},
	archivePrefix = {arXiv},
	eprint = {astro-ph/0105401},
	primaryClass = {astro-ph},
	adsurl = {https://ui.adsabs.harvard.edu/abs/2001AJ....122.1238S}
}

@ARTICLE{2008MNRAS.389.1179L,
	author = {{Lintott}, Chris J. and {Schawinski}, Kevin and {Slosar}, An{\v{z}}e and {Land}, Kate and {Bamford}, Steven and {Thomas}, Daniel and {Raddick}, M. Jordan and {Nichol}, Robert C. and {Szalay}, Alex and {Andreescu}, Dan and {Murray}, Phil and {Vandenberg}, Jan},
	title = "{Galaxy Zoo: morphologies derived from visual inspection of galaxies from the Sloan Digital Sky Survey}",
	journal = {\mnras},
	year = 2008,
	month = sep,
	volume = {389},
	number = {3},
	pages = {1179-1189},
	doi = {10.1111/j.1365-2966.2008.13689.x},
	archivePrefix = {arXiv},
	eprint = {0804.4483},
	primaryClass = {astro-ph},
	adsurl = {https://ui.adsabs.harvard.edu/abs/2008MNRAS.389.1179L}
}

@ARTICLE{2013MNRAS.435.2835W,
	author = {{Willett}, Kyle W. and {Lintott}, Chris J. and {Bamford}, Steven P. and {Masters}, Karen L. and {Simmons}, Brooke D. and {Casteels}, Kevin R.~V. and {Edmondson}, Edward M. and {Fortson}, Lucy F. and {Kaviraj}, Sugata and {Keel}, William C. and {Melvin}, Thomas and {Nichol}, Robert C. and {Raddick}, M. Jordan and {Schawinski}, Kevin and {Simpson}, Robert J. and {Skibba}, Ramin A. and {Smith}, Arfon M. and {Thomas}, Daniel},
	title = "{Galaxy Zoo 2: detailed morphological classifications for 304 122 galaxies from the Sloan Digital Sky Survey}",
	journal = {\mnras},
	year = 2013,
	month = nov,
	volume = {435},
	number = {4},
	pages = {2835-2860},
	doi = {10.1093/mnras/stt1458},
	archivePrefix = {arXiv},
	eprint = {1308.3496},
	primaryClass = {astro-ph.CO},
	adsurl = {https://ui.adsabs.harvard.edu/abs/2013MNRAS.435.2835W}
}

@ARTICLE{2018MNRAS.476.3661D,
	author = {{Dom{\'\i}nguez S{\'a}nchez}, H. and {Huertas-Company}, M. and {Bernardi}, M. and {Tuccillo}, D. and {Fischer}, J.~L.},
	title = "{Improving galaxy morphologies for SDSS with Deep Learning}",
	journal = {\mnras},
	year = 2018,
	month = feb,
	volume = {476},
	number = {3},
	pages = {3661-3676},
	doi = {10.1093/mnras/sty338},
	archivePrefix = {arXiv},
	eprint = {1711.05744},
	primaryClass = {astro-ph.GA},
	adsurl = {https://ui.adsabs.harvard.edu/abs/2018MNRAS.476.3661D}
}

@ARTICLE{2010ApJS..186..427N,
	author = {{Nair}, Preethi B. and {Abraham}, Roberto G.},
	title = "{A Catalog of Detailed Visual Morphological Classifications for 14,034 Galaxies in the Sloan Digital Sky Survey}",
	journal = {\apjs},
	year = 2010,
	month = feb,
	volume = {186},
	number = {2},
	pages = {427-456},
	doi = {10.1088/0067-0049/186/2/427},
	archivePrefix = {arXiv},
	eprint = {1001.2401},
	primaryClass = {astro-ph.CO},
	adsurl = {https://ui.adsabs.harvard.edu/abs/2010ApJS..186..427N}
}

@ARTICLE{1963ApJS....8...31D,
	author = {{de Vaucouleurs}, G.},
	title = "{Revised Classification of 1500 Bright Galaxies.}",
	journal = {\apjs},
	year = 1963,
	month = apr,
	volume = {8},
	pages = {31},
	doi = {10.1086/190084},
	adsurl = {https://ui.adsabs.harvard.edu/abs/1963ApJS....8...31D}
}

@ARTICLE{2018MNRAS.479.4509R,
	author = {{Rafieferantsoa}, Mika and {Andrianomena}, Sambatra and {Dav{\'e}}, Romeel},
	title = "{Predicting the neutral hydrogen content of galaxies from optical data using machine learning}",
	journal = {\mnras},
	year = 2018,
	month = oct,
	volume = {479},
	number = {4},
	pages = {4509-4525},
	doi = {10.1093/mnras/sty1777},
	archivePrefix = {arXiv},
	eprint = {1803.08334},
	primaryClass = {astro-ph.GA},
	adsurl = {https://ui.adsabs.harvard.edu/abs/2018MNRAS.479.4509R}
}

@ARTICLE{2011JMLR...12.2825P,
	author = {{Pedregosa}, Fabian and {Varoquaux}, Ga{\"e}l and {Gramfort}, Alexandre and {Michel}, Vincent and {Thirion}, Bertrand and {Grisel}, Olivier and {Blondel}, Mathieu and {M{\"u}ller}, Andreas and {Nothman}, Joel and {Louppe}, Gilles and {Prettenhofer}, Peter and {Weiss}, Ron and {Dubourg}, Vincent and {Vanderplas}, Jake and {Passos}, Alexandre and {Cournapeau}, David and {Brucher}, Matthieu and {Perrot}, Matthieu and {Duchesnay}, {\'E}douard},
	title = "{Scikit-learn: Machine Learning in Python}",
	journal = {Journal of Machine Learning Research},
	year = 2011,
	month = oct,
	volume = {12},
	pages = {2825-2830},
	doi = {10.48550/arXiv.1201.0490},
	archivePrefix = {arXiv},
	eprint = {1201.0490},
	primaryClass = {cs.LG},
	adsurl = {https://ui.adsabs.harvard.edu/abs/2011JMLR...12.2825P}
}

@ARTICLE{2001MachL..45....5B,
	author = {{Breiman}, Leo},
	title = "{Random Forests.}",
	journal = {Machine Learning},
	year = 2001,
	month = jan,
	volume = {45},
	pages = {5-32},
	doi = {10.1023/A:1010933404324},
	adsurl = {https://ui.adsabs.harvard.edu/abs/2001MachL..45....5B}
}

@ARTICLE{2019AJ....157...16R,
	author = {{Reis}, Itamar and {Baron}, Dalya and {Shahaf}, Sahar},
	title = "{Probabilistic Random Forest: A Machine Learning Algorithm for Noisy Data Sets}",
	journal = {\aj},
	year = 2019,
	month = jan,
	volume = {157},
	number = {1},
	eid = {16},
	pages = {16},
	doi = {10.3847/1538-3881/aaf101},
	archivePrefix = {arXiv},
	eprint = {1811.05994},
	primaryClass = {astro-ph.IM},
	adsurl = {https://ui.adsabs.harvard.edu/abs/2019AJ....157...16R}
}

@ARTICLE{2016MNRAS.460.2143W,
	author = {{Wang}, Jing and {Koribalski}, B{\"a}rbel S. and {Serra}, Paolo and {van der Hulst}, Thijs and {Roychowdhury}, Sambit and {Kamphuis}, Peter and {Chengalur}, Jayaram N.},
	title = "{New lessons from the H I size-mass relation of galaxies}",
	journal = {\mnras},
	year = 2016,
	month = aug,
	volume = {460},
	number = {2},
	pages = {2143-2151},
	doi = {10.1093/mnras/stw1099},
	archivePrefix = {arXiv},
	eprint = {1605.01489},
	primaryClass = {astro-ph.GA},
	adsurl = {https://ui.adsabs.harvard.edu/abs/2016MNRAS.460.2143W}
}

@ARTICLE{2017AJ....154...28B,
	author = {{Blanton}, Michael R. and {Bershady}, Matthew A. and {Abolfathi}, Bela and {Albareti}, Franco D. and {Allende Prieto}, Carlos and {Almeida}, Andres and {Alonso-Garc{\'\i}a}, Javier and {Anders}, Friedrich and {Anderson}, Scott F. and {Andrews}, Brett and {Aquino-Ort{\'\i}z}, Erik and {Arag{\'o}n-Salamanca}, Alfonso and {Argudo-Fern{\'a}ndez}, Maria and {Armengaud}, Eric and {Aubourg}, Eric and {Avila-Reese}, Vladimir and {Badenes}, Carles and {Bailey}, Stephen and {Barger}, Kathleen A. and {Barrera-Ballesteros}, Jorge and {Bartosz}, Curtis and {Bates}, Dominic and {Baumgarten}, Falk and {Bautista}, Julian and {Beaton}, Rachael and {Beers}, Timothy C. and {Belfiore}, Francesco and {Bender}, Chad F. and {Berlind}, Andreas A. and {Bernardi}, Mariangela and {Beutler}, Florian and {Bird}, Jonathan C. and {Bizyaev}, Dmitry and {Blanc}, Guillermo A. and {Blomqvist}, Michael and {Bolton}, Adam S. and {Boquien}, M{\'e}d{\'e}ric and {Borissova}, Jura and {van den Bosch}, Remco and {Bovy}, Jo and {Brandt}, William N. and {Brinkmann}, Jonathan and {Brownstein}, Joel R. and {Bundy}, Kevin and {Burgasser}, Adam J. and {Burtin}, Etienne and {Busca}, Nicol{\'a}s G. and {Cappellari}, Michele and {Delgado Carigi}, Maria Leticia and {Carlberg}, Joleen K. and {Carnero Rosell}, Aurelio and {Carrera}, Ricardo and {Chanover}, Nancy J. and {Cherinka}, Brian and {Cheung}, Edmond and {G{\'o}mez Maqueo Chew}, Yilen and {Chiappini}, Cristina and {Choi}, Peter Doohyun and {Chojnowski}, Drew and {Chuang}, Chia-Hsun and {Chung}, Haeun and {Cirolini}, Rafael Fernando and {Clerc}, Nicolas and {Cohen}, Roger E. and {Comparat}, Johan and {da Costa}, Luiz and {Cousinou}, Marie-Claude and {Covey}, Kevin and {Crane}, Jeffrey D. and {Croft}, Rupert A.~C. and {Cruz-Gonzalez}, Irene and {Garrido Cuadra}, Daniel and {Cunha}, Katia and {Damke}, Guillermo J. and {Darling}, Jeremy and {Davies}, Roger and {Dawson}, Kyle and {de la Macorra}, Axel and {Dell'Agli}, Flavia and {De Lee}, Nathan and {Delubac}, Timoth{\'e}e and {Di Mille}, Francesco and {Diamond-Stanic}, Aleks and {Cano-D{\'\i}az}, Mariana and {Donor}, John and {Downes}, Juan Jos{\'e} and {Drory}, Niv and {du Mas des Bourboux}, H{\'e}lion and {Duckworth}, Christopher J. and {Dwelly}, Tom and {Dyer}, Jamie and {Ebelke}, Garrett and {Eigenbrot}, Arthur D. and {Eisenstein}, Daniel J. and {Emsellem}, Eric and {Eracleous}, Mike and {Escoffier}, Stephanie and {Evans}, Michael L. and {Fan}, Xiaohui and {Fern{\'a}ndez-Alvar}, Emma and {Fernandez-Trincado}, J.~G. and {Feuillet}, Diane K. and {Finoguenov}, Alexis and {Fleming}, Scott W. and {Font-Ribera}, Andreu and {Fredrickson}, Alexander and {Freischlad}, Gordon and {Frinchaboy}, Peter M. and {Fuentes}, Carla E. and {Galbany}, Llu{\'\i}s and {Garcia-Dias}, R. and {Garc{\'\i}a-Hern{\'a}ndez}, D.~A. and {Gaulme}, Patrick and {Geisler}, Doug and {Gelfand}, Joseph D. and {Gil-Mar{\'\i}n}, H{\'e}ctor and {Gillespie}, Bruce A. and {Goddard}, Daniel and {Gonzalez-Perez}, Violeta and {Grabowski}, Kathleen and {Green}, Paul J. and {Grier}, Catherine J. and {Gunn}, James E. and {Guo}, Hong and {Guy}, Julien and {Hagen}, Alex and {Hahn}, ChangHoon and {Hall}, Matthew and {Harding}, Paul and {Hasselquist}, Sten and {Hawley}, Suzanne L. and {Hearty}, Fred and {Gonzalez Hern{\'a}ndez}, Jonay I. and {Ho}, Shirley and {Hogg}, David W. and {Holley-Bockelmann}, Kelly and {Holtzman}, Jon A. and {Holzer}, Parker H. and {Huehnerhoff}, Joseph and {Hutchinson}, Timothy A. and {Hwang}, Ho Seong and {Ibarra-Medel}, H{\'e}ctor J. and {da Silva Ilha}, Gabriele and {Ivans}, Inese I. and {Ivory}, KeShawn and {Jackson}, Kelly and {Jensen}, Trey W. and {Johnson}, Jennifer A. and {Jones}, Amy and {J{\"o}nsson}, Henrik and {Jullo}, Eric and {Kamble}, Vikrant and {Kinemuchi}, Karen and {Kirkby}, David and {Kitaura}, Francisco-Shu and {Klaene}, Mark and {Knapp}, Gillian R. and {Kneib}, Jean-Paul and {Kollmeier}, Juna A. and {Lacerna}, Ivan and {Lane}, Richard R. and {Lang}, Dustin and {Law}, David R. and {Lazarz}, Daniel and {Lee}, Youngbae and {Le Goff}, Jean-Marc and {Liang}, Fu-Heng and {Li}, Cheng and {Li}, Hongyu and {Lian}, Jianhui and {Lima}, Marcos and {Lin}, Lihwai and {Lin}, Yen-Ting and {Bertran de Lis}, Sara and {Liu}, Chao and {de Icaza Lizaola}, Miguel Angel C. and {Long}, Dan and {Lucatello}, Sara and {Lundgren}, Britt and {MacDonald}, Nicholas K. and {Deconto Machado}, Alice and {MacLeod}, Chelsea L. and {Mahadevan}, Suvrath and {Geimba Maia}, Marcio Antonio and {Maiolino}, Roberto and {Majewski}, Steven R. and {Malanushenko}, Elena and {Malanushenko}, Viktor and {Manchado}, Arturo and {Mao}, Shude and {Maraston}, Claudia and {Marques-Chaves}, Rui and {Masseron}, Thomas and {Masters}, Karen L. and {McBride}, Cameron K. and {McDermid}, Richard M. and {McGrath}, Brianne and {McGreer}, Ian D. and {Medina Pe{\~n}a}, Nicol{\'a}s and {Melendez}, Matthew and {Merloni}, Andrea and {Merrifield}, Michael R. and {Meszaros}, Szabolcs and {Meza}, Andres and {Minchev}, Ivan and {Minniti}, Dante and {Miyaji}, Takamitsu and {More}, Surhud and {Mulchaey}, John and {M{\"u}ller-S{\'a}nchez}, Francisco and {Muna}, Demitri and {Munoz}, Ricardo R. and {Myers}, Adam D. and {Nair}, Preethi and {Nandra}, Kirpal and {Correa do Nascimento}, Janaina and {Negrete}, Alenka and {Ness}, Melissa and {Newman}, Jeffrey A. and {Nichol}, Robert C. and {Nidever}, David L. and {Nitschelm}, Christian and {Ntelis}, Pierros and {O'Connell}, Julia E. and {Oelkers}, Ryan J. and {Oravetz}, Audrey and {Oravetz}, Daniel and {Pace}, Zach and {Padilla}, Nelson and {Palanque-Delabrouille}, Nathalie and {Alonso Palicio}, Pedro and {Pan}, Kaike and {Parejko}, John K. and {Parikh}, Taniya and {P{\^a}ris}, Isabelle and {Park}, Changbom and {Patten}, Alim Y. and {Peirani}, Sebastien and {Pellejero-Ibanez}, Marcos and {Penny}, Samantha and {Percival}, Will J. and {Perez-Fournon}, Ismael and {Petitjean}, Patrick and {Pieri}, Matthew M. and {Pinsonneault}, Marc and {Pisani}, Alice and {Poleski}, Rados{\l}aw and {Prada}, Francisco and {Prakash}, Abhishek and {Queiroz}, Anna B{\'a}rbara de Andrade and {Raddick}, M. Jordan and {Raichoor}, Anand and {Barboza Rembold}, Sandro and {Richstein}, Hannah and {Riffel}, Rogemar A. and {Riffel}, Rog{\'e}rio and {Rix}, Hans-Walter and {Robin}, Annie C. and {Rockosi}, Constance M. and {Rodr{\'\i}guez-Torres}, Sergio and {Roman-Lopes}, A. and {Rom{\'a}n-Z{\'u}{\~n}iga}, Carlos and {Rosado}, Margarita and {Ross}, Ashley J. and {Rossi}, Graziano and {Ruan}, John and {Ruggeri}, Rossana and {Rykoff}, Eli S. and {Salazar-Albornoz}, Salvador and {Salvato}, Mara and {S{\'a}nchez}, Ariel G. and {Aguado}, D.~S. and {S{\'a}nchez-Gallego}, Jos{\'e} R. and {Santana}, Felipe A. and {Santiago}, Bas{\'\i}lio Xavier and {Sayres}, Conor and {Schiavon}, Ricardo P. and {da Silva Schimoia}, Jaderson and {Schlafly}, Edward F. and {Schlegel}, David J. and {Schneider}, Donald P. and {Schultheis}, Mathias and {Schuster}, William J. and {Schwope}, Axel and {Seo}, Hee-Jong and {Shao}, Zhengyi and {Shen}, Shiyin and {Shetrone}, Matthew and {Shull}, Michael and {Simon}, Joshua D. and {Skinner}, Danielle and {Skrutskie}, M.~F. and {Slosar}, An{\v{z}}e and {Smith}, Verne V. and {Sobeck}, Jennifer S. and {Sobreira}, Flavia and {Somers}, Garrett and {Souto}, Diogo and {Stark}, David V. and {Stassun}, Keivan and {Stauffer}, Fritz and {Steinmetz}, Matthias and {Storchi-Bergmann}, Thaisa and {Streblyanska}, Alina and {Stringfellow}, Guy S. and {Su{\'a}rez}, Genaro and {Sun}, Jing and {Suzuki}, Nao and {Szigeti}, Laszlo and {Taghizadeh-Popp}, Manuchehr and {Tang}, Baitian and {Tao}, Charling and {Tayar}, Jamie and {Tembe}, Mita and {Teske}, Johanna and {Thakar}, Aniruddha R. and {Thomas}, Daniel and {Thompson}, Benjamin A. and {Tinker}, Jeremy L. and {Tissera}, Patricia and {Tojeiro}, Rita and {Hernandez Toledo}, Hector and {de la Torre}, Sylvain and {Tremonti}, Christy and {Troup}, Nicholas W. and {Valenzuela}, Octavio and {Martinez Valpuesta}, Inma and {Vargas-Gonz{\'a}lez}, Jaime and {Vargas-Maga{\~n}a}, Mariana and {Vazquez}, Jose Alberto and {Villanova}, Sandro and {Vivek}, M. and {Vogt}, Nicole and {Wake}, David and {Walterbos}, Rene and {Wang}, Yuting and {Weaver}, Benjamin Alan and {Weijmans}, Anne-Marie and {Weinberg}, David H. and {Westfall}, Kyle B. and {Whelan}, David G. and {Wild}, Vivienne and {Wilson}, John and {Wood-Vasey}, W.~M. and {Wylezalek}, Dominika and {Xiao}, Ting and {Yan}, Renbin and {Yang}, Meng and {Ybarra}, Jason E. and {Y{\`e}che}, Christophe and {Zakamska}, Nadia and {Zamora}, Olga and {Zarrouk}, Pauline and {Zasowski}, Gail and {Zhang}, Kai and {Zhao}, Gong-Bo and {Zheng}, Zheng and {Zheng}, Zheng and {Zhou}, Xu and {Zhou}, Zhi-Min and {Zhu}, Guangtun B. and {Zoccali}, Manuela and {Zou}, Hu},
	title = "{Sloan Digital Sky Survey IV: Mapping the Milky Way, Nearby Galaxies, and the Distant Universe}",
	journal = {\aj},
	year = 2017,
	month = jul,
	volume = {154},
	number = {1},
	eid = {28},
	pages = {28},
	doi = {10.3847/1538-3881/aa7567},
	archivePrefix = {arXiv},
	eprint = {1703.00052},
	primaryClass = {astro-ph.GA},
	adsurl = {https://ui.adsabs.harvard.edu/abs/2017AJ....154...28B}
}

@ARTICLE{1963BAAA....6...41S,
	author = {{S{\'e}rsic}, J.~L.},
	title = "{Influence of the atmospheric and instrumental dispersion on the brightness distribution in a galaxy}",
	journal = {Boletin de la Asociacion Argentina de Astronomia La Plata Argentina},
	year = 1963,
	month = feb,
	volume = {6},
	pages = {41-43},
	adsurl = {https://ui.adsabs.harvard.edu/abs/1963BAAA....6...41S}
}

@ARTICLE{1948AnAp...11..247D,
	author = {{de Vaucouleurs}, Gerard},
	title = "{Recherches sur les Nebuleuses Extragalactiques}",
	journal = {Annales d'Astrophysique},
	year = 1948,
	month = jan,
	volume = {11},
	pages = {247},
	adsurl = {https://ui.adsabs.harvard.edu/abs/1948AnAp...11..247D}
}

@ARTICLE{2010MNRAS.403..683C,
	author = {{Catinella}, Barbara and {Schiminovich}, David and {Kauffmann}, Guinevere and {Fabello}, Silvia and {Wang}, Jing and {Hummels}, Cameron and {Lemonias}, Jenna and {Moran}, Sean M. and {Wu}, Ronin and {Giovanelli}, Riccardo and {Haynes}, Martha P. and {Heckman}, Timothy M. and {Basu-Zych}, Antara R. and {Blanton}, Michael R. and {Brinchmann}, Jarle and {Budav{\'a}ri}, Tam{\'a}s and {Gon{\c{c}}alves}, Thiago and {Johnson}, Benjamin D. and {Kennicutt}, Robert C. and {Madore}, Barry F. and {Martin}, Christopher D. and {Rich}, Michael R. and {Tacconi}, Linda J. and {Thilker}, David A. and {Wild}, Vivienne and {Wyder}, Ted K.},
	title = "{The GALEX Arecibo SDSS Survey - I. Gas fraction scaling relations of massive galaxies and first data release}",
	journal = {\mnras},
	year = 2010,
	month = apr,
	volume = {403},
	number = {2},
	pages = {683-708},
	doi = {10.1111/j.1365-2966.2009.16180.x},
	archivePrefix = {arXiv},
	eprint = {0912.1610},
	primaryClass = {astro-ph.CO},
	adsurl = {https://ui.adsabs.harvard.edu/abs/2010MNRAS.403..683C}
}

@ARTICLE{2018MNRAS.476..875C,
	author = {{Catinella}, Barbara and {Saintonge}, Am{\'e}lie and {Janowiecki}, Steven and {Cortese}, Luca and {Dav{\'e}}, Romeel and {Lemonias}, Jenna J. and {Cooper}, Andrew P. and {Schiminovich}, David and {Hummels}, Cameron B. and {Fabello}, Silvia and {Ger{\'e}b}, Katinka and {Kilborn}, Virginia and {Wang}, Jing},
	title = "{xGASS: total cold gas scaling relations and molecular-to-atomic gas ratios of galaxies in the local Universe}",
	journal = {\mnras},
	year = 2018,
	month = may,
	volume = {476},
	number = {1},
	pages = {875-895},
	doi = {10.1093/mnras/sty089},
	archivePrefix = {arXiv},
	eprint = {1802.02373},
	primaryClass = {astro-ph.GA},
	adsurl = {https://ui.adsabs.harvard.edu/abs/2018MNRAS.476..875C}
}

@ARTICLE{2022ARA&A..60..319S,
	author = {{Saintonge}, Am{\'e}lie and {Catinella}, Barbara},
	title = "{The Cold Interstellar Medium of Galaxies in the Local Universe}",
	journal = {\araa},
	year = 2022,
	month = aug,
	volume = {60},
	pages = {319-361},
	doi = {10.1146/annurev-astro-021022-043545},
	archivePrefix = {arXiv},
	eprint = {2202.00690},
	primaryClass = {astro-ph.GA},
	adsurl = {https://ui.adsabs.harvard.edu/abs/2022ARA&A..60..319S}
}

@ARTICLE{2017ApJS..233...22S,
	author = {{Saintonge}, Am{\'e}lie and {Catinella}, Barbara and {Tacconi}, Linda J. and {Kauffmann}, Guinevere and {Genzel}, Reinhard and {Cortese}, Luca and {Dav{\'e}}, Romeel and {Fletcher}, Thomas J. and {Graci{\'a}-Carpio}, Javier and {Kramer}, Carsten and {Heckman}, Timothy M. and {Janowiecki}, Steven and {Lutz}, Katharina and {Rosario}, David and {Schiminovich}, David and {Schuster}, Karl and {Wang}, Jing and {Wuyts}, Stijn and {Borthakur}, Sanchayeeta and {Lamperti}, Isabella and {Roberts-Borsani}, Guido W.},
	title = "{xCOLD GASS: The Complete IRAM 30 m Legacy Survey of Molecular Gas for Galaxy Evolution Studies}",
	journal = {\apjs},
	year = 2017,
	month = dec,
	volume = {233},
	number = {2},
	eid = {22},
	pages = {22},
	doi = {10.3847/1538-4365/aa97e0},
	archivePrefix = {arXiv},
	eprint = {1710.02157},
	primaryClass = {astro-ph.GA},
	adsurl = {https://ui.adsabs.harvard.edu/abs/2017ApJS..233...22S}
}

@ARTICLE{2001ApJ...548...97S,
	author = {{Solanes}, Jos{\'e} M. and {Manrique}, Alberto and {Garc{\'\i}a-G{\'o}mez}, Carlos and {Gonz{\'a}lez-Casado}, Guillermo and {Giovanelli}, Riccardo and {Haynes}, Martha P.},
	title = "{The H I Content of Spirals. II. Gas Deficiency in Cluster Galaxies}",
	journal = {\apj},
	year = 2001,
	month = feb,
	volume = {548},
	number = {1},
	pages = {97-113},
	doi = {10.1086/318672},
	archivePrefix = {arXiv},
	eprint = {astro-ph/0007402},
	primaryClass = {astro-ph},
	adsurl = {https://ui.adsabs.harvard.edu/abs/2001ApJ...548...97S}
}

@ARTICLE{2016MNRAS.455.1294D,
	author = {{D{\'e}nes}, H. and {Kilborn}, V.~A. and {Koribalski}, B.~S. and {Wong}, O.~I.},
	title = "{H I-deficient galaxies in intermediate-density environments}",
	journal = {\mnras},
	year = 2016,
	month = jan,
	volume = {455},
	number = {2},
	pages = {1294-1308},
	doi = {10.1093/mnras/stv2391},
	archivePrefix = {arXiv},
	eprint = {1510.05343},
	primaryClass = {astro-ph.GA},
	adsurl = {https://ui.adsabs.harvard.edu/abs/2016MNRAS.455.1294D}
}

@ARTICLE{2020MNRAS.499.3233R,
	author = {{Reynolds}, T.~N. and {Westmeier}, T. and {Staveley-Smith}, L.},
	title = "{H I deficiencies and asymmetries in HIPASS galaxies}",
	journal = {\mnras},
	year = 2020,
	month = dec,
	volume = {499},
	number = {3},
	pages = {3233-3242},
	doi = {10.1093/mnras/staa3126},
	archivePrefix = {arXiv},
	eprint = {2010.03720},
	primaryClass = {astro-ph.GA},
	adsurl = {https://ui.adsabs.harvard.edu/abs/2020MNRAS.499.3233R}
}

@ARTICLE{2013MNRAS.436...34C,
	author = {{Catinella}, Barbara and {Schiminovich}, David and {Cortese}, Luca and {Fabello}, Silvia and {Hummels}, Cameron B. and {Moran}, Sean M. and {Lemonias}, Jenna J. and {Cooper}, Andrew P. and {Wu}, Ronin and {Heckman}, Timothy M. and {Wang}, Jing},
	title = "{The GALEX Arecibo SDSS Survey - VIII. Final data release. The effect of group environment on the gas content of massive galaxies}",
	journal = {\mnras},
	year = 2013,
	month = nov,
	volume = {436},
	number = {1},
	pages = {34-70},
	doi = {10.1093/mnras/stt1417},
	archivePrefix = {arXiv},
	eprint = {1308.1676},
	primaryClass = {astro-ph.CO},
	adsurl = {https://ui.adsabs.harvard.edu/abs/2013MNRAS.436...34C}
}

@ARTICLE{2013ApJ...772..119L,
       author = {{Lilly}, Simon J. and {Carollo}, C. Marcella and {Pipino}, Antonio and {Renzini}, Alvio and {Peng}, Yingjie},
        title = "{Gas Regulation of Galaxies: The Evolution of the Cosmic Specific Star Formation Rate, the Metallicity-Mass-Star-formation Rate Relation, and the Stellar Content of Halos}",
      journal = {\apj},
         year = 2013,
        month = aug,
       volume = {772},
       number = {2},
          eid = {119},
        pages = {119},
          doi = {10.1088/0004-637X/772/2/119},
archivePrefix = {arXiv},
       eprint = {1303.5059},
 primaryClass = {astro-ph.CO},
       adsurl = {https://ui.adsabs.harvard.edu/abs/2013ApJ...772..119L}
}

@ARTICLE{2020MNRAS.496.3531G,
       author = {{Gogate}, A.~R. and {Verheijen}, M.~A.~W. and {Deshev}, B.~Z. and {van Gorkom}, J.~H. and {Montero-Casta{\~n}o}, M. and {van der Hulst}, J.~M. and {Jaff{\'e}}, Y.~L. and {Poggianti}, B.~M.},
        title = "{BUDHIES IV: Deep 21-cm neutral Hydrogen, optical, and UV imaging data of Abell 963 and Abell 2192 at z $\simeq$ 0.2}",
      journal = {\mnras},
         year = 2020,
        month = aug,
       volume = {496},
       number = {3},
        pages = {3531-3552},
          doi = {10.1093/mnras/staa1680},
archivePrefix = {arXiv},
       eprint = {2302.12197},
 primaryClass = {astro-ph.GA},
       adsurl = {https://ui.adsabs.harvard.edu/abs/2020MNRAS.496.3531G}
}

@ARTICLE{2019MNRAS.490...96S,
	author = {{Stevens}, Adam R.~H. and {Diemer}, Benedikt and {Lagos}, Claudia del P. and {Nelson}, Dylan and {Obreschkow}, Danail and {Wang}, Jing and {Marinacci}, Federico},
	title = "{Origin of the galaxy H I size-mass relation}",
	journal = {\mnras},
	year = 2019,
	month = nov,
	volume = {490},
	number = {1},
	pages = {96-113},
	doi = {10.1093/mnras/stz2513},
	archivePrefix = {arXiv},
	eprint = {1908.11149},
	primaryClass = {astro-ph.GA},
	adsurl = {https://ui.adsabs.harvard.edu/abs/2019MNRAS.490...96S}
}

@ARTICLE{2011MNRAS.418.1587T,
	author = {{Taylor}, Edward N. and {Hopkins}, Andrew M. and {Baldry}, Ivan K. and {Brown}, Michael J.~I. and {Driver}, Simon P. and {Kelvin}, Lee S. and {Hill}, David T. and {Robotham}, Aaron S.~G. and {Bland-Hawthorn}, Joss and {Jones}, D.~H. and {Sharp}, R.~G. and {Thomas}, Daniel and {Liske}, Jochen and {Loveday}, Jon and {Norberg}, Peder and {Peacock}, J.~A. and {Bamford}, Steven P. and {Brough}, Sarah and {Colless}, Matthew and {Cameron}, Ewan and {Conselice}, Christopher J. and {Croom}, Scott M. and {Frenk}, C.~S. and {Gunawardhana}, Madusha and {Kuijken}, Konrad and {Nichol}, R.~C. and {Parkinson}, H.~R. and {Phillipps}, S. and {Pimbblet}, K.~A. and {Popescu}, C.~C. and {Prescott}, Matthew and {Sutherland}, W.~J. and {Tuffs}, R.~J. and {van Kampen}, Eelco and {Wijesinghe}, D.},
	title = "{Galaxy And Mass Assembly (GAMA): stellar mass estimates}",
	journal = {\mnras},
	year = 2011,
	month = dec,
	volume = {418},
	number = {3},
	pages = {1587-1620},
	doi = {10.1111/j.1365-2966.2011.19536.x},
	archivePrefix = {arXiv},
	eprint = {1108.0635},
	primaryClass = {astro-ph.CO},
	adsurl = {https://ui.adsabs.harvard.edu/abs/2011MNRAS.418.1587T}
}

@ARTICLE{2009MNRAS.393.1324B,
	author = {{Bamford}, Steven P. and {Nichol}, Robert C. and {Baldry}, Ivan K. and {Land}, Kate and {Lintott}, Chris J. and {Schawinski}, Kevin and {Slosar}, An{\v{z}}e and {Szalay}, Alexander S. and {Thomas}, Daniel and {Torki}, Mehri and {Andreescu}, Dan and {Edmondson}, Edward M. and {Miller}, Christopher J. and {Murray}, Phil and {Raddick}, M. Jordan and {Vandenberg}, Jan},
	title = "{Galaxy Zoo: the dependence of morphology and colour on environment*}",
	journal = {\mnras},
	year = 2009,
	month = mar,
	volume = {393},
	number = {4},
	pages = {1324-1352},
	doi = {10.1111/j.1365-2966.2008.14252.x},
	archivePrefix = {arXiv},
	eprint = {0805.2612},
	primaryClass = {astro-ph},
	adsurl = {https://ui.adsabs.harvard.edu/abs/2009MNRAS.393.1324B}
}

@ARTICLE{2007AJ....133..734B,
	author = {{Blanton}, Michael R. and {Roweis}, Sam},
	title = "{K-Corrections and Filter Transformations in the Ultraviolet, Optical, and Near-Infrared}",
	journal = {\aj},
	year = 2007,
	month = feb,
	volume = {133},
	number = {2},
	pages = {734-754},
	doi = {10.1086/510127},
	archivePrefix = {arXiv},
	eprint = {astro-ph/0606170},
	primaryClass = {astro-ph},
	adsurl = {https://ui.adsabs.harvard.edu/abs/2007AJ....133..734B}
}

@ARTICLE{1976ApJS...32..409T,
	author = {{Turner}, E.~L. and {Gott}, III, J.~R.},
	title = "{Groups of galaxies. I. A catalog.}",
	journal = {\apjs},
	year = 1976,
	month = nov,
	volume = {32},
	pages = {409-427},
	doi = {10.1086/190403},
	adsurl = {https://ui.adsabs.harvard.edu/abs/1976ApJS...32..409T}
}

@ARTICLE{1982ApJ...257..423H,
	author = {{Huchra}, J.~P. and {Geller}, M.~J.},
	title = "{Groups of Galaxies. I. Nearby groups}",
	journal = {\apj},
	year = 1982,
	month = jun,
	volume = {257},
	pages = {423-437},
	doi = {10.1086/160000},
	adsurl = {https://ui.adsabs.harvard.edu/abs/1982ApJ...257..423H}
}

@ARTICLE{1982Natur.300..407Z,
	author = {{Zeldovich}, Ia. B. and {Einasto}, J. and {Shandarin}, S.~F.},
	title = "{Giant voids in the Universe}",
	journal = {\nat},
	year = 1982,
	month = dec,
	volume = {300},
	number = {5891},
	pages = {407-413},
	doi = {10.1038/300407a0},
	adsurl = {https://ui.adsabs.harvard.edu/abs/1982Natur.300..407Z}
}

@ARTICLE{1981PASP...93....5B,
	author = {{Baldwin}, J.~A. and {Phillips}, M.~M. and {Terlevich}, R.},
	title = "{Classification parameters for the emission-line spectra of extragalactic objects.}",
	journal = {\pasp},
	year = 1981,
	month = feb,
	volume = {93},
	pages = {5-19},
	doi = {10.1086/130766},
	adsurl = {https://ui.adsabs.harvard.edu/abs/1981PASP...93....5B}
}

@ARTICLE{2008MNRAS.383.1519C,
       author = {{Cortese}, L. and {Minchin}, R.~F. and {Auld}, R.~R. and {Davies}, J.~I. and {Catinella}, B. and {Momjian}, E. and {Rosenberg}, J.~L. and {Taylor}, R. and {Gavazzi}, G. and {O'Neil}, K. and {Baes}, M. and {Boselli}, A. and {Bothun}, G. and {Koribalski}, B. and {Schneider}, S. and {van Driel}, W.},
        title = "{The Arecibo Galaxy Environment Survey - II. A HI view of the Abell cluster 1367 and its outskirts}",
      journal = {\mnras},
         year = 2008,
        month = feb,
       volume = {383},
       number = {4},
        pages = {1519-1537},
          doi = {10.1111/j.1365-2966.2007.12664.x},
archivePrefix = {arXiv},
       eprint = {0711.0684},
 primaryClass = {astro-ph},
       adsurl = {https://ui.adsabs.harvard.edu/abs/2008MNRAS.383.1519C}
}

@ARTICLE{2018A_A...609A..17J,
       author = {{Jones}, M.~G. and {Espada}, D. and {Verdes-Montenegro}, L. and {Huchtmeier}, W.~K. and {Lisenfeld}, U. and {Leon}, S. and {Sulentic}, J. and {Sabater}, J. and {Jones}, D.~E. and {Sanchez}, S. and {Garrido}, J.},
        title = "{The AMIGA sample of isolated galaxies. XIII. The HI content of an almost ``nurture free'' sample}",
      journal = {\aap},
         year = 2018,
        month = jan,
       volume = {609},
          eid = {A17},
        pages = {A17},
          doi = {10.1051/0004-6361/201731448},
archivePrefix = {arXiv},
       eprint = {1710.03034},
 primaryClass = {astro-ph.GA},
       adsurl = {https://ui.adsabs.harvard.edu/abs/2018A&A...609A..17J}
}

@ARTICLE{2020MNRAS.494.1114S,
       author = {{Steyrleithner}, P. and {Hensler}, G. and {Boselli}, A.},
        title = "{The effect of ram-pressure stripping on dwarf galaxies}",
      journal = {\mnras},
         year = 2020,
        month = may,
       volume = {494},
       number = {1},
        pages = {1114-1127},
          doi = {10.1093/mnras/staa775},
archivePrefix = {arXiv},
       eprint = {2003.09591},
 primaryClass = {astro-ph.GA},
       adsurl = {https://ui.adsabs.harvard.edu/abs/2020MNRAS.494.1114S}
}

@ARTICLE{2003AJ....125.2975L,
       author = {{Lee}, Henry and {McCall}, Marshall L. and {Richer}, Michael G.},
        title = "{Uncovering Additional Clues to Galaxy Evolution. II. The Environmental Impact of the Virgo Cluster on the Evolution of Dwarf Irregular Galaxies}",
      journal = {\aj},
         year = 2003,
        month = jun,
       volume = {125},
       number = {6},
        pages = {2975-2997},
          doi = {10.1086/375304},
archivePrefix = {arXiv},
       eprint = {astro-ph/0303359},
 primaryClass = {astro-ph},
       adsurl = {https://ui.adsabs.harvard.edu/abs/2003AJ....125.2975L}
}

@ARTICLE{2016MNRAS.455.2323M,
       author = {{Mistani}, Pouria A. and {Sales}, Laura V. and {Pillepich}, Annalisa and {Sanchez-Janssen}, Rub{\'e}n and {Vogelsberger}, Mark and {Nelson}, Dylan and {Rodriguez-Gomez}, Vicente and {Torrey}, Paul and {Hernquist}, Lars},
        title = "{On the assembly of dwarf galaxies in clusters and their efficient formation of globular clusters}",
      journal = {\mnras},
         year = 2016,
        month = jan,
       volume = {455},
       number = {3},
        pages = {2323-2336},
          doi = {10.1093/mnras/stv2435},
archivePrefix = {arXiv},
       eprint = {1509.00030},
 primaryClass = {astro-ph.GA},
       adsurl = {https://ui.adsabs.harvard.edu/abs/2016MNRAS.455.2323M}
}

@ARTICLE{2015Natur.521..192P,
       author = {{Peng}, Y. and {Maiolino}, R. and {Cochrane}, R.},
        title = "{Strangulation as the primary mechanism for shutting down star formation in galaxies}",
      journal = {\nat},
         year = 2015,
        month = may,
       volume = {521},
       number = {7551},
        pages = {192-195},
          doi = {10.1038/nature14439},
archivePrefix = {arXiv},
       eprint = {1505.03143},
 primaryClass = {astro-ph.GA},
       adsurl = {https://ui.adsabs.harvard.edu/abs/2015Natur.521..192P}
}

@ARTICLE{2022A&A...657A...9C,
       author = {{Castignani}, G. and {Combes}, F. and {Jablonka}, P. and {Finn}, R.~A. and {Rudnick}, G. and {Vulcani}, B. and {Desai}, V. and {Zaritsky}, D. and {Salom{\'e}}, P.},
        title = "{Virgo filaments. I. Processing of gas in cosmological filaments around the Virgo cluster}",
      journal = {\aap},
         year = 2022,
        month = jan,
       volume = {657},
          eid = {A9},
        pages = {A9},
          doi = {10.1051/0004-6361/202040141},
archivePrefix = {arXiv},
       eprint = {2101.04389},
 primaryClass = {astro-ph.GA},
       adsurl = {https://ui.adsabs.harvard.edu/abs/2022A&A...657A...9C}
}

@ARTICLE{2012MNRAS.423..787T,
       author = {{Taylor}, R. and {Davies}, J.~I. and {Auld}, R. and {Minchin}, R.~F.},
        title = "{The Arecibo Galaxy Environment Survey - V. The Virgo cluster (I)}",
      journal = {\mnras},
         year = 2012,
        month = jun,
       volume = {423},
       number = {1},
        pages = {787-810},
          doi = {10.1111/j.1365-2966.2012.20914.x},
archivePrefix = {arXiv},
       eprint = {1203.3094},
 primaryClass = {astro-ph.GA},
       adsurl = {https://ui.adsabs.harvard.edu/abs/2012MNRAS.423..787T}
}

@ARTICLE{2018ApJ...852..142C,
       author = {{Crone Odekon}, Mary and {Hallenbeck}, Gregory and {Haynes}, Martha P. and {Koopmann}, Rebecca A. and {Phi}, An and {Wolfe}, Pierre-Francois},
        title = "{The Effect of Filaments and Tendrils on the H I Content of Galaxies}",
      journal = {\apj},
         year = 2018,
        month = jan,
       volume = {852},
       number = {2},
          eid = {142},
        pages = {142},
          doi = {10.3847/1538-4357/aaa1e8},
archivePrefix = {arXiv},
       eprint = {1712.05045},
 primaryClass = {astro-ph.GA},
       adsurl = {https://ui.adsabs.harvard.edu/abs/2018ApJ...852..142C}
}

@ARTICLE{2005AJ....130.2598G,
       author = {{Giovanelli}, Riccardo and {Haynes}, Martha P. and {Kent}, Brian R. and {Perillat}, Philip and {Saintonge}, Amelie and {Brosch}, Noah and {Catinella}, Barbara and {Hoffman}, G. Lyle and {Stierwalt}, Sabrina and {Spekkens}, Kristine and {Lerner}, Mikael S. and {Masters}, Karen L. and {Momjian}, Emmanuel and {Rosenberg}, Jessica L. and {Springob}, Christopher M. and {Boselli}, Alessandro and {Charmandaris}, Vassilis and {Darling}, Jeremy K. and {Davies}, Jonathan and {Garcia Lambas}, Diego and {Gavazzi}, Giuseppe and {Giovanardi}, Carlo and {Hardy}, Eduardo and {Hunt}, Leslie K. and {Iovino}, Angela and {Karachentsev}, Igor D. and {Karachentseva}, Valentina E. and {Koopmann}, Rebecca A. and {Marinoni}, Christian and {Minchin}, Robert and {Muller}, Erik and {Putman}, Mary and {Pantoja}, Carmen and {Salzer}, John J. and {Scodeggio}, Marco and {Skillman}, Evan and {Solanes}, Jose M. and {Valotto}, Carlos and {van Driel}, Wim and {van Zee}, Liese},
        title = "{The Arecibo Legacy Fast ALFA Survey. I. Science Goals, Survey Design, and Strategy}",
      journal = {\aj},
         year = 2005,
        month = dec,
       volume = {130},
       number = {6},
        pages = {2598-2612},
          doi = {10.1086/497431},
archivePrefix = {arXiv},
       eprint = {astro-ph/0508301},
 primaryClass = {astro-ph},
       adsurl = {https://ui.adsabs.harvard.edu/abs/2005AJ....130.2598G}
}

@ARTICLE{2000AJ....120.1579Y,
       author = {{York}, Donald G. and {Adelman}, J. and {Anderson}, Jr., John E. and {Anderson}, Scott F. and {Annis}, James and {Bahcall}, Neta A. and {Bakken}, J.~A. and {Barkhouser}, Robert and {Bastian}, Steven and {Berman}, Eileen and {Boroski}, William N. and {Bracker}, Steve and {Briegel}, Charlie and {Briggs}, John W. and {Brinkmann}, J. and {Brunner}, Robert and {Burles}, Scott and {Carey}, Larry and {Carr}, Michael A. and {Castander}, Francisco J. and {Chen}, Bing and {Colestock}, Patrick L. and {Connolly}, A.~J. and {Crocker}, J.~H. and {Csabai}, Istv{\'a}n and {Czarapata}, Paul C. and {Davis}, John Eric and {Doi}, Mamoru and {Dombeck}, Tom and {Eisenstein}, Daniel and {Ellman}, Nancy and {Elms}, Brian R. and {Evans}, Michael L. and {Fan}, Xiaohui and {Federwitz}, Glenn R. and {Fiscelli}, Larry and {Friedman}, Scott and {Frieman}, Joshua A. and {Fukugita}, Masataka and {Gillespie}, Bruce and {Gunn}, James E. and {Gurbani}, Vijay K. and {de Haas}, Ernst and {Haldeman}, Merle and {Harris}, Frederick H. and {Hayes}, J. and {Heckman}, Timothy M. and {Hennessy}, G.~S. and {Hindsley}, Robert B. and {Holm}, Scott and {Holmgren}, Donald J. and {Huang}, Chi-hao and {Hull}, Charles and {Husby}, Don and {Ichikawa}, Shin-Ichi and {Ichikawa}, Takashi and {Ivezi{\'c}}, {\v{Z}}eljko and {Kent}, Stephen and {Kim}, Rita S.~J. and {Kinney}, E. and {Klaene}, Mark and {Kleinman}, A.~N. and {Kleinman}, S. and {Knapp}, G.~R. and {Korienek}, John and {Kron}, Richard G. and {Kunszt}, Peter Z. and {Lamb}, D.~Q. and {Lee}, B. and {Leger}, R. French and {Limmongkol}, Siriluk and {Lindenmeyer}, Carl and {Long}, Daniel C. and {Loomis}, Craig and {Loveday}, Jon and {Lucinio}, Rich and {Lupton}, Robert H. and {MacKinnon}, Bryan and {Mannery}, Edward J. and {Mantsch}, P.~M. and {Margon}, Bruce and {McGehee}, Peregrine and {McKay}, Timothy A. and {Meiksin}, Avery and {Merelli}, Aronne and {Monet}, David G. and {Munn}, Jeffrey A. and {Narayanan}, Vijay K. and {Nash}, Thomas and {Neilsen}, Eric and {Neswold}, Rich and {Newberg}, Heidi Jo and {Nichol}, R.~C. and {Nicinski}, Tom and {Nonino}, Mario and {Okada}, Norio and {Okamura}, Sadanori and {Ostriker}, Jeremiah P. and {Owen}, Russell and {Pauls}, A. George and {Peoples}, John and {Peterson}, R.~L. and {Petravick}, Donald and {Pier}, Jeffrey R. and {Pope}, Adrian and {Pordes}, Ruth and {Prosapio}, Angela and {Rechenmacher}, Ron and {Quinn}, Thomas R. and {Richards}, Gordon T. and {Richmond}, Michael W. and {Rivetta}, Claudio H. and {Rockosi}, Constance M. and {Ruthmansdorfer}, Kurt and {Sandford}, Dale and {Schlegel}, David J. and {Schneider}, Donald P. and {Sekiguchi}, Maki and {Sergey}, Gary and {Shimasaku}, Kazuhiro and {Siegmund}, Walter A. and {Smee}, Stephen and {Smith}, J. Allyn and {Snedden}, S. and {Stone}, R. and {Stoughton}, Chris and {Strauss}, Michael A. and {Stubbs}, Christopher and {SubbaRao}, Mark and {Szalay}, Alexander S. and {Szapudi}, Istvan and {Szokoly}, Gyula P. and {Thakar}, Anirudda R. and {Tremonti}, Christy and {Tucker}, Douglas L. and {Uomoto}, Alan and {Vanden Berk}, Dan and {Vogeley}, Michael S. and {Waddell}, Patrick and {Wang}, Shu-i. and {Watanabe}, Masaru and {Weinberg}, David H. and {Yanny}, Brian and {Yasuda}, Naoki and {SDSS Collaboration}},
        title = "{The Sloan Digital Sky Survey: Technical Summary}",
      journal = {\aj},
         year = 2000,
        month = sep,
       volume = {120},
       number = {3},
        pages = {1579-1587},
          doi = {10.1086/301513},
archivePrefix = {arXiv},
       eprint = {astro-ph/0006396},
 primaryClass = {astro-ph},
       adsurl = {https://ui.adsabs.harvard.edu/abs/2000AJ....120.1579Y}
}

@ARTICLE{1976ApJ...209L...1P,
       author = {{Petrosian}, V.},
        title = "{Surface Brightness and Evolution of Galaxies}",
      journal = {\apjl},
         year = 1976,
        month = dec,
       volume = {210},
        pages = {L53},
          doi = {10.1086/18230110.1086/182253},
       adsurl = {https://ui.adsabs.harvard.edu/abs/1976ApJ...209L...1P}
}

@ARTICLE{2001AJ....121.2358B,
       author = {{Blanton}, Michael R. and {Dalcanton}, Julianne and {Eisenstein}, Daniel and {Loveday}, Jon and {Strauss}, Michael A. and {SubbaRao}, Mark and {Weinberg}, David H. and {Anderson}, Jr., John E. and {Annis}, James and {Bahcall}, Neta A. and {Bernardi}, Mariangela and {Brinkmann}, J. and {Brunner}, Robert J. and {Burles}, Scott and {Carey}, Larry and {Castander}, Francisco J. and {Connolly}, Andrew J. and {Csabai}, Istv{\'a}n and {Doi}, Mamoru and {Finkbeiner}, Douglas and {Friedman}, Scott and {Frieman}, Joshua A. and {Fukugita}, Masataka and {Gunn}, James E. and {Hennessy}, G.~S. and {Hindsley}, Robert B. and {Hogg}, David W. and {Ichikawa}, Takashi and {Ivezi{\'c}}, {\v{Z}}eljko and {Kent}, Stephen and {Knapp}, G.~R. and {Lamb}, D.~Q. and {Leger}, R. French and {Long}, Daniel C. and {Lupton}, Robert H. and {McKay}, Timothy A. and {Meiksin}, Avery and {Merelli}, Aronne and {Munn}, Jeffrey A. and {Narayanan}, Vijay and {Newcomb}, Matt and {Nichol}, R.~C. and {Okamura}, Sadanori and {Owen}, Russell and {Pier}, Jeffrey R. and {Pope}, Adrian and {Postman}, Marc and {Quinn}, Thomas and {Rockosi}, Constance M. and {Schlegel}, David J. and {Schneider}, Donald P. and {Shimasaku}, Kazuhiro and {Siegmund}, Walter A. and {Smee}, Stephen and {Snir}, Yehuda and {Stoughton}, Chris and {Stubbs}, Christopher and {Szalay}, Alexander S. and {Szokoly}, Gyula P. and {Thakar}, Aniruddha R. and {Tremonti}, Christy and {Tucker}, Douglas L. and {Uomoto}, Alan and {Vanden Berk}, Dan and {Vogeley}, Michael S. and {Waddell}, Patrick and {Yanny}, Brian and {Yasuda}, Naoki and {York}, Donald G.},
        title = "{The Luminosity Function of Galaxies in SDSS Commissioning Data}",
      journal = {\aj},
         year = 2001,
        month = may,
       volume = {121},
       number = {5},
        pages = {2358-2380},
          doi = {10.1086/320405},
archivePrefix = {arXiv},
       eprint = {astro-ph/0012085},
 primaryClass = {astro-ph},
       adsurl = {https://ui.adsabs.harvard.edu/abs/2001AJ....121.2358B}
}

@ARTICLE{2011AJ....142..170H,
       author = {{Haynes}, Martha P. and {Giovanelli}, Riccardo and {Martin}, Ann M. and {Hess}, Kelley M. and {Saintonge}, Am{\'e}lie and {Adams}, Elizabeth A.~K. and {Hallenbeck}, Gregory and {Hoffman}, G. Lyle and {Huang}, Shan and {Kent}, Brian R. and {Koopmann}, Rebecca A. and {Papastergis}, Emmanouil and {Stierwalt}, Sabrina and {Balonek}, Thomas J. and {Craig}, David W. and {Higdon}, Sarah J.~U. and {Kornreich}, David A. and {Miller}, Jeffrey R. and {O'Donoghue}, Aileen A. and {Olowin}, Ronald P. and {Rosenberg}, Jessica L. and {Spekkens}, Kristine and {Troischt}, Parker and {Wilcots}, Eric M.},
        title = "{The Arecibo Legacy Fast ALFA Survey: The {\ensuremath{\alpha}}.40 H I Source Catalog, Its Characteristics and Their Impact on the Derivation of the H I Mass Function}",
      journal = {\aj},
         year = 2011,
        month = nov,
       volume = {142},
       number = {5},
          eid = {170},
        pages = {170},
          doi = {10.1088/0004-6256/142/5/170},
archivePrefix = {arXiv},
       eprint = {1109.0027},
 primaryClass = {astro-ph.CO},
       adsurl = {https://ui.adsabs.harvard.edu/abs/2011AJ....142..170H}
}

@ARTICLE{2025A&A...696A.113T,
       author = {{Taylor}, Rhys},
        title = "{Quantifying the completeness and reliability of visual source extraction: An examination of eight thousand data cubes by eye}",
      journal = {\aap},
         year = 2025,
        month = apr,
       volume = {696},
          eid = {A113},
        pages = {A113},
          doi = {10.1051/0004-6361/202451606},
archivePrefix = {arXiv},
       eprint = {2503.07430},
 primaryClass = {astro-ph.IM},
       adsurl = {https://ui.adsabs.harvard.edu/abs/2025A&A...696A.113T}
}

@ARTICLE{2011AstBu..66....1K,
       author = {{Karachentsev}, I.~D. and {Makarov}, D.~I. and {Karachentseva}, V.~E. and {Melnyk}, O.~V.},
        title = "{Catalog of nearby isolated galaxies in the volume z < 0.01}",
      journal = {Astrophysical Bulletin},
         year = 2011,
        month = jan,
       volume = {66},
       number = {1},
          eid = {1},
        pages = {1},
          doi = {10.1134/S1990341311010019},
archivePrefix = {arXiv},
       eprint = {1103.3990},
 primaryClass = {astro-ph.CO},
       adsurl = {https://ui.adsabs.harvard.edu/abs/2011AstBu..66....1K}
}

@ARTICLE{2011MNRAS.417..370G,
       author = {{Guo}, Quan and {Cole}, Shaun and {Eke}, Vincent and {Frenk}, Carlos},
        title = "{The satellite luminosity functions of galaxies in Sloan Digital Sky Survey}",
      journal = {\mnras},
         year = 2011,
        month = oct,
       volume = {417},
       number = {1},
        pages = {370-381},
          doi = {10.1111/j.1365-2966.2011.19270.x},
archivePrefix = {arXiv},
       eprint = {1101.2674},
 primaryClass = {astro-ph.CO},
       adsurl = {https://ui.adsabs.harvard.edu/abs/2011MNRAS.417..370G}
}

@ARTICLE{2012MNRAS.424.2574W,
       author = {{Wang}, Wenting and {White}, Simon D.~M.},
        title = "{Satellite abundances around bright isolated galaxies}",
      journal = {\mnras},
         year = 2012,
        month = aug,
       volume = {424},
       number = {4},
        pages = {2574-2598},
          doi = {10.1111/j.1365-2966.2012.21256.x},
archivePrefix = {arXiv},
       eprint = {1203.0009},
 primaryClass = {astro-ph.CO},
       adsurl = {https://ui.adsabs.harvard.edu/abs/2012MNRAS.424.2574W}
}

@ARTICLE{1996ApJ...461..609S,
       author = {{Solanes}, Jose M. and {Giovanelli}, Riccardo and {Haynes}, Martha P.},
        title = "{The H i Content of Spirals. I. Field Galaxy H i Mass Functions and H i Mass--Optical Size Regressions}",
      journal = {\apj},
         year = 1996,
        month = apr,
       volume = {461},
        pages = {609},
          doi = {10.1086/177089},
archivePrefix = {arXiv},
       eprint = {astro-ph/9511003},
 primaryClass = {astro-ph},
       adsurl = {https://ui.adsabs.harvard.edu/abs/1996ApJ...461..609S}
}

@ARTICLE{2014MNRAS.444..667D,
       author = {{D{\'e}nes}, H. and {Kilborn}, V.~A. and {Koribalski}, B.~S.},
        title = "{New H I scaling relations to probe the H I content of galaxies via global H I-deficiency maps}",
      journal = {\mnras},
         year = 2014,
        month = oct,
       volume = {444},
       number = {1},
        pages = {667-681},
          doi = {10.1093/mnras/stu1337},
archivePrefix = {arXiv},
       eprint = {1407.1619},
 primaryClass = {astro-ph.GA},
       adsurl = {https://ui.adsabs.harvard.edu/abs/2014MNRAS.444..667D}
}

@ARTICLE{2011ApJ...732...93T,
       author = {{Toribio}, M. Carmen and {Solanes}, Jos{\'e} M. and {Giovanelli}, Riccardo and {Haynes}, Martha P. and {Martin}, Ann M.},
        title = "{H I Content and Optical Properties of Field Galaxies from the ALFALFA Survey. II. Multivariate Analysis of a Galaxy Sample in Low-density Environments}",
      journal = {\apj},
         year = 2011,
        month = may,
       volume = {732},
       number = {2},
          eid = {93},
        pages = {93},
          doi = {10.1088/0004-637X/732/2/93},
archivePrefix = {arXiv},
       eprint = {1103.0990},
 primaryClass = {astro-ph.CO},
       adsurl = {https://ui.adsabs.harvard.edu/abs/2011ApJ...732...93T}
}

@ARTICLE{Hearin_2023,
      author = {{Hearin}, Andrew P. and {Chaves-Montero}, Jon{\'a}s and {Alarcon}, Alex and {Becker}, Matthew R. and {Benson}, Andrew},
        title = "{DSPS: Differentiable stellar population synthesis}",
      journal = {MNRAS},
        year = 2023,
        month = may,
      volume = {521},
      number = {2},
        pages = {1741-1756},
          doi = {10.1093/mnras/stad456},
archivePrefix = {arXiv},
      eprint = {2112.06830},
primaryClass = {astro-ph.GA},
      adsurl = {https://ui.adsabs.harvard.edu/abs/2023MNRAS.521.1741H}
}

@ARTICLE{Kuchner_2017,
       author = {{Kuchner}, U. and {Ziegler}, B. and {Verdugo}, M. and {Bamford}, S. and {H{\"a}u{\ss}ler}, B.},
        title = "{The effects of the cluster environment on the galaxy mass-size relation in MACS J1206.2-0847}",
      journal = {\aap},
         year = 2017,
        month = aug,
       volume = {604},
          eid = {A54},
        pages = {A54},
          doi = {10.1051/0004-6361/201630252},
archivePrefix = {arXiv},
       eprint = {1705.03839},
 primaryClass = {astro-ph.GA},
       adsurl = {https://ui.adsabs.harvard.edu/abs/2017A&A...604A..54K}
}

@ARTICLE{Kelvin_2012,
       author = {{Kelvin}, Lee S. and {Driver}, Simon P. and {Robotham}, Aaron S.~G. and {Hill}, David T. and {Alpaslan}, Mehmet and {Baldry}, Ivan K. and {Bamford}, Steven P. and {Bland-Hawthorn}, Joss and {Brough}, Sarah and {Graham}, Alister W. and {H{\"a}ussler}, Boris and {Hopkins}, Andrew M. and {Liske}, Jochen and {Loveday}, Jon and {Norberg}, Peder and {Phillipps}, Steven and {Popescu}, Cristina C. and {Prescott}, Matthew and {Taylor}, Edward N. and {Tuffs}, Richard J.},
        title = "{Galaxy And Mass Assembly (GAMA): Structural Investigation of Galaxies via Model Analysis}",
      journal = {\mnras},
         year = 2012,
        month = apr,
       volume = {421},
       number = {2},
        pages = {1007-1039},
          doi = {10.1111/j.1365-2966.2012.20355.x},
archivePrefix = {arXiv},
       eprint = {1112.1956},
 primaryClass = {astro-ph.CO},
       adsurl = {https://ui.adsabs.harvard.edu/abs/2012MNRAS.421.1007K}
}

@ARTICLE{Genel_2018,
       author = {{Genel}, Shy and {Nelson}, Dylan and {Pillepich}, Annalisa and {Springel}, Volker and {Pakmor}, R{\"u}diger and {Weinberger}, Rainer and {Hernquist}, Lars and {Naiman}, Jill and {Vogelsberger}, Mark and {Marinacci}, Federico and {Torrey}, Paul},
        title = "{The size evolution of star-forming and quenched galaxies in the IllustrisTNG simulation}",
      journal = {\mnras},
         year = 2018,
        month = mar,
       volume = {474},
       number = {3},
        pages = {3976-3996},
          doi = {10.1093/mnras/stx3078},
archivePrefix = {arXiv},
       eprint = {1707.05327},
 primaryClass = {astro-ph.GA},
       adsurl = {https://ui.adsabs.harvard.edu/abs/2018MNRAS.474.3976G}
}

@ARTICLE{Larson1980,
   author = {{Larson}, R.~B. and {Tinsley}, B.~M. and {Caldwell}, C.~N.},
    title = "{The evolution of disk galaxies and the origin of S0 galaxies}",
  journal = {\apj},
     year = 1980,
    month = may,
   volume = 237,
    pages = {692-707},
      doi = {10.1086/157917},
   adsurl = {http://adsabs.harvard.edu/abs/1980ApJ...237..692L}
}

@ARTICLE{Taylor2020,
       author = {{Taylor}, Rhys and {K{\"o}ppen}, Joachim and {J{\'a}chym}, Pavel and {Minchin}, Robert and {Palou{\v{s}}}, Jan and {W{\"u}nsch}, Richard},
        title = "{Faint and Fading Tails: The Fate of Stripped H I Gas in Virgo Cluster Galaxies}",
      journal = {\aj},
         year = 2020,
        month = may,
       volume = {159},
       number = {5},
          eid = {218},
        pages = {218},
          doi = {10.3847/1538-3881/ab6988},
archivePrefix = {arXiv},
       eprint = {2001.03385},
 primaryClass = {astro-ph.GA},
       adsurl = {https://ui.adsabs.harvard.edu/abs/2020AJ....159..218T}
}

@ARTICLE{2017MNRAS.464.3796T,
       author = {{Teimoorinia}, Hossen and {Ellison}, Sara L. and {Patton}, David R.},
        title = "{Pattern recognition in the ALFALFA.70 and Sloan Digital Sky Surveys: a catalogue of {\ensuremath{\sim}}500 000 H I gas fraction estimates based on artificial neural networks}",
      journal = {\mnras},
         year = 2017,
        month = feb,
       volume = {464},
       number = {4},
        pages = {3796-3811},
          doi = {10.1093/mnras/stw2606},
archivePrefix = {arXiv},
       eprint = {1610.02341},
 primaryClass = {astro-ph.GA},
       adsurl = {https://ui.adsabs.harvard.edu/abs/2017MNRAS.464.3796T}
}

@ARTICLE{2020ApJ...900..142W,
       author = {{Wu}, John F.},
        title = "{Connecting Optical Morphology, Environment, and H I Mass Fraction for Low-redshift Galaxies Using Deep Learning}",
      journal = {\apj},
         year = 2020,
        month = sep,
       volume = {900},
       number = {2},
          eid = {142},
        pages = {142},
          doi = {10.3847/1538-4357/abacbb},
archivePrefix = {arXiv},
       eprint = {2001.00018},
 primaryClass = {astro-ph.GA},
       adsurl = {https://ui.adsabs.harvard.edu/abs/2020ApJ...900..142W}
}

@dataset{catalog,
  author       = {Janák, Filip},
  title        = {Catalog of expected (predicted) neutral hydrogen
                   masses for a sample of non-isolated galaxies
                  },
  month        = apr,
  year         = 2026,
  publisher    = {Zenodo},
  doi          = {10.5281/zenodo.19711065},
  url          = {https://doi.org/10.5281/zenodo.19711065},
}

@misc{appendix_figures,
  author       = {Janák, Filip and
                  Deshev, Boris and
                  Nagy, Roman and
                  Taylor, Rhys},
  title        = {A machine learning approach to estimating HI
                   deficiency in galaxies; additional plots
                  },
  month        = apr,
  year         = 2026,
  publisher    = {Zenodo},
  doi          = {10.5281/zenodo.19819331},
  url          = {https://doi.org/10.5281/zenodo.19819331},
}

\appendix

\section{Comparison between used and discarded ALFALFA-SDSS samples}
\label{Appendix_0}

In Fig. A.1 \citep{appendix_figures} we compare the distributions of the input features and the target variable (\Mhi) for the ALFALFA-SDSS sample used in this work (i.e. galaxies cross-matched with the \cite{2017A&A...602A.100T} environment catalog; $\sim$15\,000 objects) and the discarded sample ($\sim$14\,000 objects). Because the \cite{2017A&A...602A.100T} catalog is based on spectroscopic targets from SDSS, it prefers more luminous galaxies, as spectroscopy is generally unavailable for fainter systems. Consequently, the discarded sample extends to lower luminosities (reaching absolute magnitudes of about $-7$), whereas the used sample does not include galaxies fainter than approximately $-10$. This luminosity bias is naturally reflected in the colors. Finally, both samples exhibit very similar distributions in all radii and concentrations, as well as in \Mhi.

\section{Feature distributions of IG sample}
\label{Appendix_1}

Here we present feature distributions of IG sample with respect to the target variable (Figs. B.1 and B.2 \citep{appendix_figures}). Neutral hydrogen shows similar trends across all bands for the three types of radii and concentration indices. Absolute magnitudes in different bands have distributions of similar shape, but are progressively shifted toward brighter galaxies with increasing wavelength. In contrast, the two colors have different distributions from each other, with more offset peaks. The target variable (decadic logarithm of \Mhi in solar units) covers the range from $ \sim 7 $ to $ \sim 10.5 $, although with very few galaxies below $ \sim 8.5 $.

Feature distributions show that colors are only very weakly related with the neutral hydrogen, while concentration indices appear entirely unrelated with \Mhi. The original motivation for including concentrations and colors as features was their role as morphology tracers (see Sect. \ref{ch2.4}). However, there is not any significant relation between morphology type and \Mhi in our IG sample (its distribution is qualitatively and quantitatively the same as the distribution of a whole sample shown in top left of Fig. \ref{morph}).

\section{Linear relation between \Mhi and the optical diameter for different morphologies}
\label{Appendix_2}

\begin{figure}
	\centering
	\includegraphics[width=0.5\textwidth]{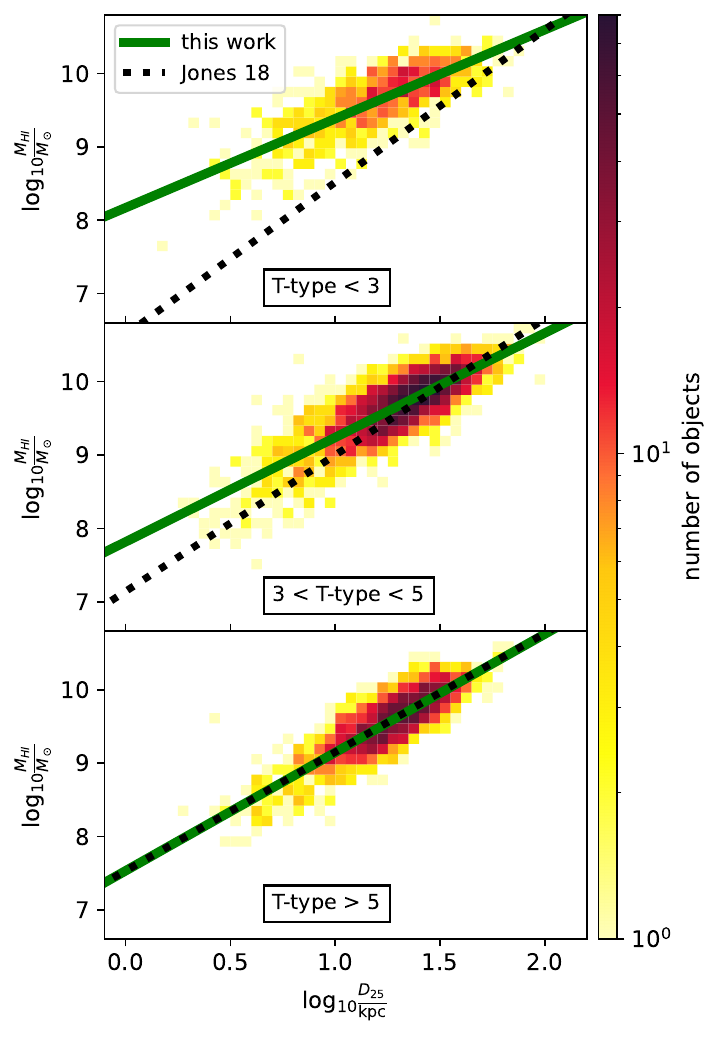}
	\caption{\hi mass for isolated galaxies versus the optical diameter $ D_{25} $ for three subsamples according to the T-type: $\mathrm{T\mbox{-}type}<3$ (851 galaxies), $3<\mathrm{T\mbox{-}type}<5$ (2956 galaxies) and $\mathrm{T\mbox{-}type}>5$ (1913 galaxies). The linear fit is plotted with the green line, compared with \cite{2018A_A...609A..17J} (black dotted line).}
	\label{lin_model_all}
\end{figure}

Here we compare the relation between the \hi mass and the optical diameter $D_{25}$ (see Sect. \ref{ch3.3}) in our sample of IG for different morphologies. We split the IG sample into three subsamples according to the T-type: $\mathrm{T\mbox{-}type}<3$ (851 galaxies), $3<\mathrm{T\mbox{-}type}<5$ (2956 galaxies) and $\mathrm{T\mbox{-}type}>5$ (1913 galaxies). The results are shown in Fig. \ref{lin_model_all} and Table \ref{abmorph_tab}, where we compare linear fits with values obtained by \cite{2018A_A...609A..17J}. We find that the parameter $a$ decreases with the T-type while the parameter $b$ increases, in contrast to \cite{2018A_A...609A..17J}, who reported the opposite trend. The largest discrepancies in $a$ and $b$ occur for ETGs, whereas for the late-type subsample ($\mathrm{T\mbox{-}type}>5$) we recover the same $a$ and $b$ values as those of \cite{2018A_A...609A..17J}. The disagreement for ETGs may reflect their limited representation in our sample, combined with the limitation of adopted conversion from $R_{90}$ to $D_{25}$, which is expected to be most reliable for LTGs (as discussed in Sect. \ref{ch3.3}).

\begin{table}
	\centering
    \scriptsize
	\caption{Parameters $a$ and $b$ from Eq. \ref{MHI_vs_D}, describing the linear relation between \Mhi and optical diameter (both on a logarithmic scale) for isolated galaxies for different morphologies, compared with values obtained by \cite{2018A_A...609A..17J}.}
	\label{abmorph_tab}
	\begin{tabular}{c|cccc}
		\hline
		T-type & $<3$ & $3-5$ & $>5$ \\
		\hline
		$a$ \citep{2018A_A...609A..17J} & $6.44\pm0.59$ & $7.14\pm0.18$ & $7.53\pm0.24$ \\
		\hline
		$a$ (this work) & $8.17\pm0.04$ & $7.81\pm0.02$ & $7.53\pm0.03$ \\
		\hline
		$b$ \citep{2018A_A...609A..17J} & $2.05\pm0.42$ & $1.86\pm0.12$ & $1.62\pm0.18$ \\
		\hline
		$b$ (this work) & $1.21\pm0.04$ & $1.42\pm0.02$ & $1.62\pm0.02$ \\
		\hline
	\end{tabular}
\end{table}

\section{Comparison between IG and nIG samples}
\label{Appendix_3}

Here we compare the distributions of features and the target variable (\Mhi) for the IG sample (6\,982 objects) and the nIG sample (8\,232 objects). Figure D.1 \citep{appendix_figures} shows histograms for IG (red, hatched) and nIG (black, line). Both samples exhibit very similar profiles, particularly in the \Mhi distribution, with only minor differences: compared to IG, nIG are shifted toward brighter magnitudes, larger optical sizes, higher concentration indices, and redder colors, indicating a shift toward ETGs.

\section{Time evolution of optical parameters}
\label{Appendix_4}
In Fig. E.1 \citep{appendix_figures} we present the four separate star formation histories and the associated evolutions of the optical parameters used to estimate the changes in the derived \hi deficiency discussed in Sect. \ref{time_evolution}.

\end{document}